\documentclass[twocolumn,showpacs,showkeys,preprintnumbers,amsmath,amssymb]{revtex4}
\usepackage{graphicx}% Include figure files
\usepackage{dcolumn}% Align table columns on decimal point
\usepackage{bm}% bold math
\begin{document}

%\preprint{APS/123-QED}
%\preprint{prePrint2}

\title{$\beta$ Decay and Isomeric Properties of Neutron-Rich Ca and 
Sc Isotopes}

\author{H. L. Crawford$^{1,2}$,
R. V. F. Janssens$^{3}$,
P. F. Mantica$^{1,2}$,
J. S. Berryman$^{1,2}$,\\
R. Broda$^{4}$,
M. P. Carpenter$^{3}$,
N. Cieplicka$^{4}$,
B. Fornal$^{4}$,
G. F. Grinyer$^{2}$,
N. Hoteling$^{3,5}$,\\
B. P. Kay$^{3}$,
T. Lauritsen$^{3}$,
K. Minamisono$^{2}$,
I. Stefanescu$^{3,5}$,
J. B. Stoker$^{1,2}$,\\
W. B. Walters$^{5}$, and
S. Zhu$^{3}$
}

\affiliation{$^{(1)}$
Department of Chemistry, Michigan State University,
East Lansing, Michigan 48824}
\affiliation{$^{(2)}$ 
National Superconducting Cyclotron
Laboratory, Michigan State University,
East Lansing, Michigan 48824} 
\affiliation{$^{(3)}$
Physics Division, Argonne National Laboratory
Argonne, Illinois 60439}
\affiliation{$^{(4)}$
Institute of Nuclear Physics, Polish Academy of Sciences
Cracow, Poland PL-31342}
\affiliation{$^{(5)}$
Department of Chemistry and Biochemistry, University of Maryland,
College Park, Maryland 20742}

\date{\today}% It is always \today, today,
             %  but any date may be explicitly specified

\begin{abstract}

The isomeric and $\beta$-decay properties of neutron-rich 
$^{53-57}$Sc and $^{53,54}$Ca nuclei near neutron number 
$N$=32 are reported, and the low-energy level schemes 
of $^{53,54,56}$Sc and $^{53-57}$Ti are presented.  The 
low-energy level structures of the $_{21}$Sc isotopes are 
discussed in terms of the coupling of the valence 
$1f_{7/2}$ proton to states in the corresponding $_{20}$Ca 
cores.  Implications with respect to the robustness of the 
$N$=32 subshell closure are discussed, as well as the 
repercussions for a possible $N$=34 subshell closure.

\end{abstract}

% PACS, the Physics and Astronomy
                             % Classification Scheme.
%Use showkeys class option if keyword
                              %display desired
\maketitle

%%%%%%%%%%%%%%%%%%%%%%%%%  INTRODUCTION   %%%%%%%%%%%%%%%%%%%%%%%%%%%%%%%

\section{\label{sec1:level1}Introduction}

The nuclear shell model~\cite{Maye55} was developed in 
part to explain the observed extra stability of nuclei 
along the valley of $\beta$ stability corresponding to 
`magic' numbers of protons and/or neutrons.  The 
magic numbers at 2, 8, 20, etc. were explained 
as corresponding to closed-shell nucleonic 
configurations.  While the shell model 
is robust near stability, there is evidence at the extremes of 
the nuclear chart that the single-particle level ordering changes, 
leading to the erosion of 
some shell closures and/or the appearance of new `magic' 
numbers in exotic nuclei.  The attractive 
proton-neutron monopole interaction~\cite{Otsu01} can 
produce drastic changes in level ordering, and in regions 
of low single-particle density, result in modified shell gaps.  
For example, the monopole shift of the $\nu$$1f_{5/2}$ 
orbit with changing $\pi$$1f_{7/2}$ occupancy 
has been invoked to explain the observed $N$=32 
subshell closure in nuclei in the neighborhood of 
the doubly-magic nucleus $^{48}_{20}$Ca.  

The $N$=32 subshell closure has been well-established 
experimentally in $_{20}$Ca~\cite{Huck85, Gade06}, 
$_{22}$Ti~\cite{Jans02} and $_{24}$Cr~\cite{Pris01} 
on the basis of the systematic behavior of the 
$E(2^{+}_{1})$ energies as a function of neutron 
number in the even-even nuclei of these isotopic 
chains.  Additional evidence for 
$N$=32 as a new subshell closure in neutron-rich nuclei 
below $Z$=28 comes from the low $B(E2:2^{+}\rightarrow0^{+})$  
transition probabilities measured for $^{54}$Ti~\cite{Dinc05} and 
$^{56}$Cr~\cite{Burg05}.  The high-spin structures of 
$^{50,52,54}$Ti~\cite{Jans02} are also indicative 
of a sizeable $N$=32 shell gap.  The separation
between the 6$_{1}^{+}$ and the closely spaced 8$_{1}^{+}$, 
9$_{1}^{+}$ and 10$_{1}^{+}$ levels in $^{54}$Ti was taken to indicate 
the substantial energy cost in promoting a $2p_{3/2}$
neutron to either of the $\nu$$2p_{1/2}$ or $\nu$$1f_{5/2}$ orbitals. 

Migration of the $\nu$$1f_{5/2}$ orbit may be such that, 
beyond $N$=32, a gap 
between the $\nu$$2p_{1/2}$ and the $\nu$$1f_{5/2}$ 
orbitals develops.  The results of shell-model calculations with the GXPF1 
effective interaction point to a 
$N$=34 subshell closure in the 
neutron-rich Ti and Ca isotopes~\cite{Honm02}.  
The prediction of a new magic 
number at $N$=34 for the Ti isotopes is inconsistent with experimental 
results in $^{56}$Ti~\cite{Lidd04_PRL, Dinc05}.  
However, shell-model calculations with the slightly-modified 
GXPF1A effective interaction~\cite{Honm05} 
suggest that a $N$=34 subshell gap may still be 
present in the Ca isotopes.

While the level structure of $^{54}$Ca remains 
difficult to reach experimentally, the possible 
development of a $N$=34 shell closure in this 
region can be examined indirectly.  Some insight into the relative 
collectivity of the $_{20}$Ca isotopes can be obtained 
from the energy separation between the $(\pi$$1f_{7/2})^{3}_{7/2-}$ 
ground states and $(\pi$$1f_{7/2})^{3}_{5/2-}$ excited states in 
the odd-mass $_{23}$V isotopes.  These cluster states were 
described in detail by Paar~\cite{Paar73}, and shown to 
depend on the interaction between the proton cluster and 
underlying core.  In the cases of $^{51}$V$_{28}$~\cite{Bunk55} and 
$^{55}$V$_{32}$~\cite{Mant03,Zhu07}, with 
three protons beyond doubly-magic $^{48}$Ca$_{28}$ and 
$^{52}$Ca$_{32}$ respectively, the separations between the ground state and 
5/2$^{-}$ cluster states are 320 and 323 keV.  
However, for $^{53}$V$_{30}$~\cite{Pron76} and $^{57}$V$_{34}$~\cite{Lidd05}, 
with three protons beyond $^{50}$Ca$_{30}$ and $^{54}$Ca$_{34}$ respectively, 
the separations are only 128 and 113 keV, 
suggesting that the $^{54}$Ca core is more like that of 
$^{50}$Ca than magic $^{48}$Ca or $^{52}$Ca.

The possible $N$=34 
subshell closure has also been probed by tracking the 
monopole migration of the $\nu$$1f_{5/2}$ orbital in neighboring 
nuclei.  Previous work by Liddick $\it{et~al.}$~\cite{Lidd04}, 
following the systematic evolution of the $\nu$$1f_{5/2}$ 
level through consideration of the ground-state spin 
and parity assignments of odd-odd V and Sc isotopes, 
indicated an inversion of the $\nu$$1f_{5/2}$ and 
$\nu$$2p_{1/2}$ single-particle orbitals in moving from 
the $_{23}$V to the $_{21}$Sc isotopes.  The
apparent close spacing of the $\nu$$1f_{5/2}$ and 
$\nu$$2p_{1/2}$ orbits in the $_{22}$Ti isotopes, leading
to the absence of the $N$=34 subshell closure for $Z$=22,
also supports the observed evolution of the neutron 
single-particle states.  Energy spacings between 
high-spin states in $^{54}$Ti and the first three yrast 
states in $^{56}$Ti~\cite{Forn04} both suggest that the 
$\nu$$1f_{5/2}$ and $\nu$$2p_{1/2}$ orbitals are nearly 
degenerate.

\begin{figure}[!tb]
\centering
\includegraphics[width = 0.48\textwidth]{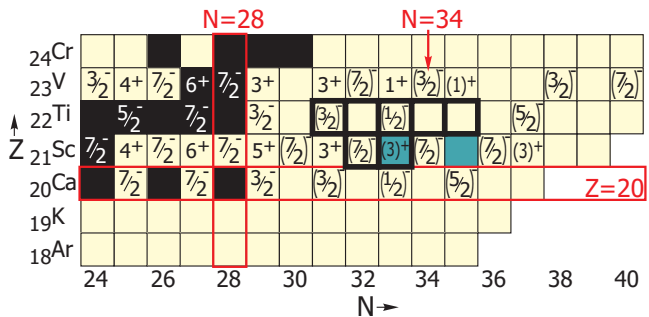}
\caption[Chart of Nuclides]{
(Color online) The region of the chart of the nuclides under 
consideration in the present work.  Black squares represent 
stable nuclei, while nuclei for which low-energy levels were 
identified in the present work from the $\beta$ 
decay of parent nuclei are highlighted by a thick black border.  
Nuclei for which $\mu$s isomeric states permitted access to 
low-energy levels are shaded.  
Ground-state spins and parities are included where known, 
and all even-even nuclides have ground state $J^{\pi}$ = 0$^{+}$.}
\label{fig:nuclides}
\end{figure}

Here, we report on the low-energy structures of 
$^{53,54}$Sc populated following the $\beta$ decay 
of $^{53,54}$Ca respectively, as well as on the 
low-energy levels of $^{54,56}$Sc populated by isomeric 
$\gamma$-ray decay.  The 
low-energy levels of the $_{21}$Sc isotopes can provide 
insight into the structure of even-even Ca isotopes, 
as these Sc levels result from the coupling of the odd proton to states 
in the Ca core.  Additionally, we report on newly observed 
states in $^{53-57}$Ti populated in the $\beta$ decay 
of $^{53-57}$Sc, respectively.  The nuclei studied in the 
present work are highlighted in Fig.\ \ref{fig:nuclides}, 
where known ground-state spins and parities are included 
for the $_{20}$Ca, $_{21}$Sc, $_{22}$Ti and $_{23}$V isotopes.  
The present results reaffirm the 
validity of the $N$=32 subshell closure, and the inversion 
of the $\nu$$1f_{5/2}$ and $\nu$$2p_{1/2}$ single particle 
levels moving from the V to the Sc isotopes.  However, the 
apparent compression of the low-energy structure of the neutron-rich 
Sc isotopes suggests that there may not be a significant 
$N$=34 gap between the $\nu$$2p_{1/2}$ and $\nu$$1f_{5/2}$ 
single-particle states in the $_{21}$Sc isotopic chain.

%%%%%%%%%%%%%%%%%%%%%%%%%%%%%  EXPERIMENTAL   %%%%%%%%%%%%%%%%%%%%%%%%%%%

\section{\label{sec2:level1}Experimental Procedure}

The $\beta$ decay and isomeric properties of the 
neutron-rich $^{53,54}$Ca and $^{53-57}$Sc parent nuclides 
were deduced in experiments carried out using the 
experimental facilities at National Superconducting 
Cyclotron Laboratory (NSCL) at Michigan State University.

A low-energy beam of $^{76}$Ge$^{12+}$ was accelerated 
to 11.6~MeV/nucleon in the K500 cyclotron, and, 
following foil stripping to produce $^{76}$Ge$^{30+}$, 
accelerated to the full energy of 130~MeV/nucleon in 
the K1200 cyclotron.  The $^{76}$Ge beam was fragmented 
in a 352-mg/cm$^{2}$ thick Be target located at the 
object position of the A1900 fragment separator~\cite{Morr03}.  
The secondary fragments of interest were selected in the 
A1900 using a 45-mg/cm$^{2}$ Al wedge and 5\% momentum 
slits located at the intermediate dispersive image of the 
separator.

Fully-stripped secondary fragments were sent to the NSCL 
Beta Counting System (BCS)\cite{Pris03} located in the S2 
experimental vault.  The central implantation detector of the BCS was a 
995-$\mu$m thick double-sided Si microstrip detector (DSSD).  
The DSSD was segmented into 40 1-mm wide strips in both 
the horizontal and vertical dimensions, for a total of 1600 
pixels.  Three Si PIN detectors, with respective thicknesses of 
297~$\mu$m, 297~$\mu$m and 488~$\mu$m, were located upstream 
of the DSSD.  Six single-sided 
Si strip detectors (SSSDs) were placed downstream of the 
DSSD.  The last two SSSDs provided a veto for light particles 
that arrived at the counting system with the secondary beam.  
The range of the fragments was adjusted with a 4-mm Al degrader located 
immediately upstream of the BCS detectors, to implant the ions 
near the center of the DSSD.

\begin{figure}[!tb]
\centering
\includegraphics[width = 0.45\textwidth]{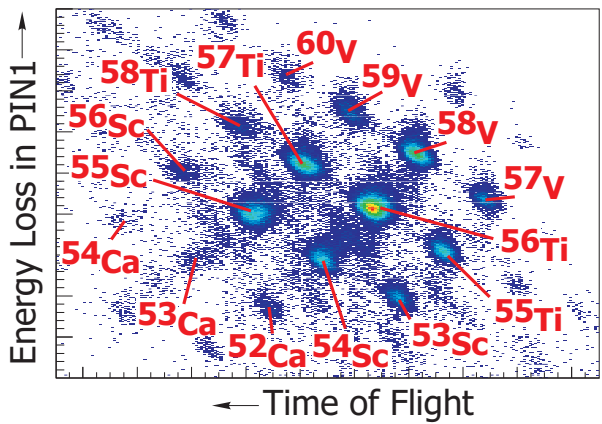}
\caption[Representative Particle ID Plot]{(Color online) Particle 
identification spectrum obtained from fragment energy loss in 
PIN1 versus time-of-flight.  A1900 magnetic rigidity settings 
of $B$$\rho_{1,2}$ = 4.103~Tm and $B$$\rho_{3,4}$ = 4.030~Tm 
were used to produce the observed cocktail beam.}
\label{fig:particleID}
\end{figure}

Implanted fragments were unambiguously identified by a combination of 
multiple energy loss signals, time-of-flight, 
and the fragment separator magnetic rigidity ($B$$\rho$) setting.  The 
fragment energy loss was measured in the three PIN 
detectors, and fragment time-of-flight was determined 
as the time difference between a particle impinging on 
a plastic scintillator detector positioned at the dispersive image 
of the A1900 separator, and the first Si PIN detector (PIN1).  
Fig.\ \ref{fig:particleID} illustrates a representative particle 
identification spectrum for the experiment investigating $^{53,54}$Ca and 
$^{53-56}$Sc.  The particle ID spectrum that included $^{57}$Sc 
was presented previously~\cite{Mant08}.  The secondary 
beam was defocused to maximize the active area of the DSSD 
that was illuminated.  The average implantation rate over the 
entire face of the DSSD was $\leq$ 100~Hz.

Fragment-$\beta$ correlations were established in 
software by requiring a high-energy implantation 
event in a single pixel ($\it{x}$,$\it{y}$) of the DSSD, 
where $\it{x}$ is a horizontal strip and $\it{y}$ is a 
vertical one.  An implantion event was one which 
had a valid PIN1 signal as well as a high-energy signal in 
a single strip of both the front and back of the DSSD.  A 
$\beta$-decay event was one with a high-gain 
signal above threshold in the front and back of the DSSD and 
no signal in PIN1. $\beta$-decay events were correlated with 
implantations only if they occured in the same pixel, within 
a specified correlation time window that was adjusted to be at least ten 
half-lives of the parent nucleus.  The differences between the absolute 
time stamps of correlated $\beta$ and implantation events were 
histogrammed to generate decay curves. 

The profile of implantations over the DSSD surface was such that 
pixels near the center of the detector experienced a significantly 
higher implantation rate than those at the detector edge.  The 
higher implantation rate led to a larger accumulated activity 
in the central pixels.  To avoid random correlations when using 
longer correlation times, a subset of pixels on the DSSD was 
defined for which the time between implantations was $\geq$100~s.  
Corresponding to pixels on the edge of the detector, this subset of 
pixels included 50\% of the detector surface, and was used for the
analysis of decays with long correlation times.  
Details of the correlation 
times and the pixels used in analysis are included with the results 
for each nucleus.  

Prompt and delayed $\gamma$-ray detection was accomplished 
using sixteen detectors from the Segmented Germanium 
Array (SeGA)~\cite{Muel01}.  The Ge detectors were oriented 
in two concentric rings of 8 detectors each, surrounding 
the beam pipe containing the BCS detectors.  One detector was 
omitted from the analysis of data for $^{53,54}$Ca and 
$^{53-56}$Sc due to poor gain stability. A photopeak efficiency for 
$\gamma$-ray detection of 7\% at 1~MeV and 3\% at 3~MeV was achieved.  
The energy resolution of all detectors was $\leq$ 3.5~keV for the 
1.3-MeV transition in $^{60}$Co.  Prompt $\gamma$ rays emitted 
following short-lived isomeric decay were detected within a  20-$\mu$s 
window following an implantation event.  A time-to-amplitude converter 
(TAC) was used to measure the time elapsed between an implantation event and 
the observation of isomeric $\gamma$ rays in any of the SeGA detectors, 
permitting measurement of $\mu$s isomer half-lives.

%%%%%%%%%%%%%%%%%%%%%%%%%%%%%%%%  RESULTS   %%%%%%%%%%%%%%%%%%%%%%%%%%%%%

\section{\label{sec3:level1}Results}

\subsection{\label{sec3:level2}$^{53}$Ca $\beta$ decay}

\begin{figure}[!tb]
\centering
\includegraphics[width = 0.4\textwidth]{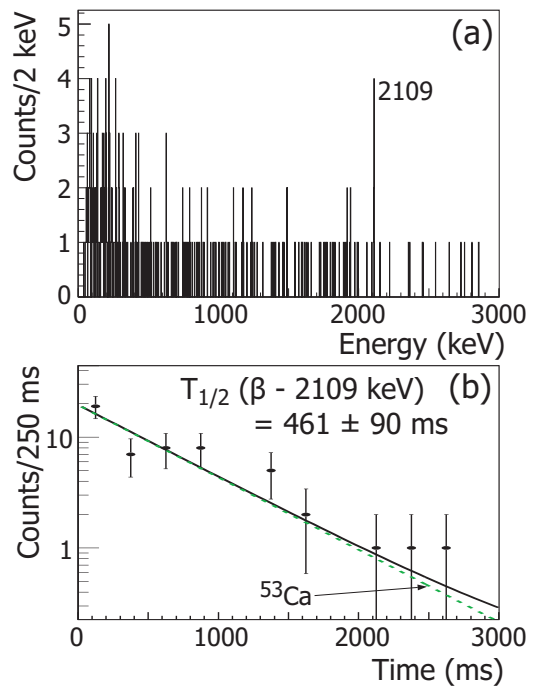}
\caption[53Ca Gamma Rays and Decay]{(a) $\beta$-delayed 
$\gamma$-ray spectrum for $^{53}$Ca in the range of 0-3~MeV.  The 
spectrum includes events within the first 5~s following a 
$^{53}$Ca implantation event, in detector pixels in which the 
average time between implantations was $\geq$ 100~s.  
The 2109-keV transition is the sole $\gamma$ ray observed.  
(b) Decay curve for $^{53}$Ca using a 5-s correlation time, considering 
all DSSD pixels, with the additional requirement of a coincident 2109-keV 
$\gamma$ ray.  The curve was fitted with a single exponential 
decay.}
\label{fig:53Ca}
\end{figure}

A correlation time of 5~s was used to analyze the $\beta$ decay of 
$^{53}$Ca to $^{53}$Sc, where the expected half-life of $^{53}$Ca was 
on the order of several hundred milliseconds.  With a relatively long 
correlation window, the analysis was restricted to pixels located 
towards the edge of the implantation DSSD.    
The $\beta$-delayed $\gamma$-ray spectrum 
for $^{53}$Ca in the range 0 to 3~MeV is presented in Fig.\ 
\ref{fig:53Ca}(a).  One transition at 2109.0$\pm$0.3~keV has been 
assigned to the $\beta$ decay of $^{53}$Ca.  The newly-assigned 
2109-keV transition was initially reported by us in Ref.~\cite{Craw09}, and 
has been recently confirmed in one-proton knockout from 
$^{54}$Ti~\cite{McDa09}.  An absolute $\gamma$ intensity of 56$\pm$12\% 
was determined for this 2109-keV transition from a Gaussian fit of 
the peak area, the absolute efficiency of SeGA, 
and the number of $\beta$ rays correlated 
with $^{53}$Ca parent decays.  The absolute intensity of this 
transition suggests that the majority of $\beta$ intensity from 
the decay of $^{53}$Ca, excluding the previously established 
40$\pm$10\% $\beta$$n$ branch~\cite{Lang83}, 
proceeds through the 2109-keV excited state.  This level 
has been tentatively assigned spin and parity 3/2$^{-}$, on the 
basis of the apparent allowed $\beta$ decay from a 
(1/2)$^{-}$ $^{53}$Ca ground state.  The only other possibility for 
a state populated in allowed $\beta$ decay is a 1/2$^{-}$ 
assignment, which is excluded due to the observed 
direct transition to the 7/2$^{-}$ $^{53}$Sc ground state.

The decay curve for $^{53}$Ca gated on the 2109-keV transition 
is presented in Fig.\ \ref{fig:53Ca}(b).  Here, all pixels of the 
DSSD were considered, since the additional $\gamma$-coincidence 
requirement significantly reduced background events.  The data were fitted 
with a single exponential decay, and a half-life value of 461$\pm$90~ms 
was extracted.  This value is higher than the value (230$\pm$60~ms) obtained 
previously by Mantica ${\it et~al.}$~\cite{Mant08}, which was deduced from 
a half-life curve with no additional $\gamma$-ray requirement.  The 
required $\gamma$ coincidence in the present result eliminates 
uncertainties arising from daughter decay and background contributions, 
and provides a more accurate half-life determination.  However, 
as noted in Ref.~\cite{Mant08}, the possibility of a second 
$\beta$-decaying state in $^{53}$Ca cannot be excluded.

\subsection{\label{sec3:level2b}$^{54}$Ca $\beta$ decay}

\begin{figure}[!tb]
\centering
\includegraphics[width = 0.4\textwidth]{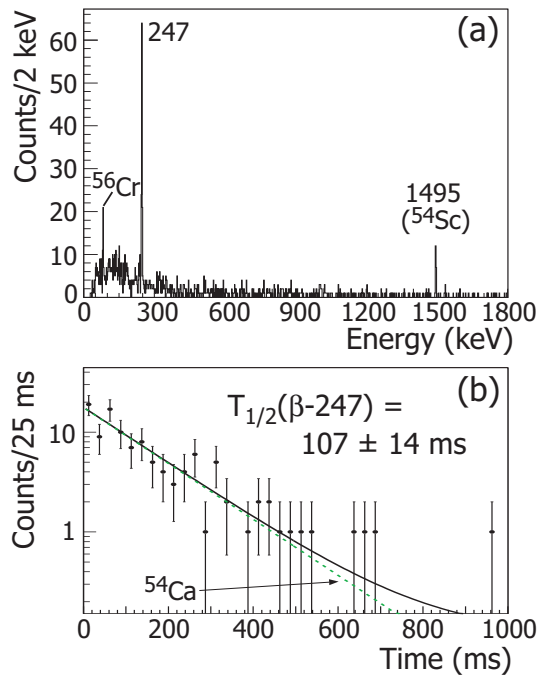}
\caption[54Ca Gamma Rays and Decay]{(Color online) (a) 
$\beta$-delayed $\gamma$-ray spectrum for $^{54}$Ca in the range of 
0-1.8~MeV.  The spectrum includes events within the first 
1~s following a $^{54}$Ca implantation.  Transitions are 
marked by their energy in keV.  (b) Decay curve for 
$^{54}$Ca-correlated $\beta$-decay events occuring within 
1~s of a $^{54}$Ca implantation, with a requirement of a 
coincident 247-keV $\gamma$ ray.  The curve was fitted with 
a single exponential decay and a constant background.}
\label{fig:54Ca}
\end{figure}

Previous measurement of the $^{54}$Ca half-life suggested a 
value of order 100~ms~\cite{Mant08}.  With such an expectation, 
a correlation time of 1~s was used for this short-lived nucleus, 
and data over the entire surface of the DSSD were considered.  
The $\beta$-delayed $\gamma$-ray spectrum of $^{54}$Ca in the 
range 0 to 1.8~MeV is 
shown in Fig.\ \ref{fig:54Ca}(a).  One transition at 
247.3$\pm$0.3~keV, with an absolute intensity of 65$\pm$9\%, 
has been assigned to the $\beta$ decay of 
$^{54}$Ca.  This transition corresponds 
to the $\gamma$ ray observed by Mantica ${\it et~al.}$~\cite{Mant08}, 
and placed as a ground state transition depopulating a 247-keV 
state in $^{54}$Sc, assigned as $J^{\pi}$ = 1$^{+}$.  
Missing $\beta$ intensity can be attributed to a possible 
$\beta$$n$ branch in the $^{54}$Ca decay.  

Fig.\ \ref{fig:54Ca}(b) provides the decay curve obtained 
by requiring a coincidence with the 247-keV $\gamma$ ray.  
This decay curve yielded 
a half-life value of 107$\pm$14~ms, within 1$\sigma$ 
of the previous value~\cite{Mant08}.

\subsection{\label{sec3:level2c}$^{53}$Sc $\beta$ decay}

\begin{figure*}[!tbp]
\centering
\includegraphics[width = 0.8\textwidth]{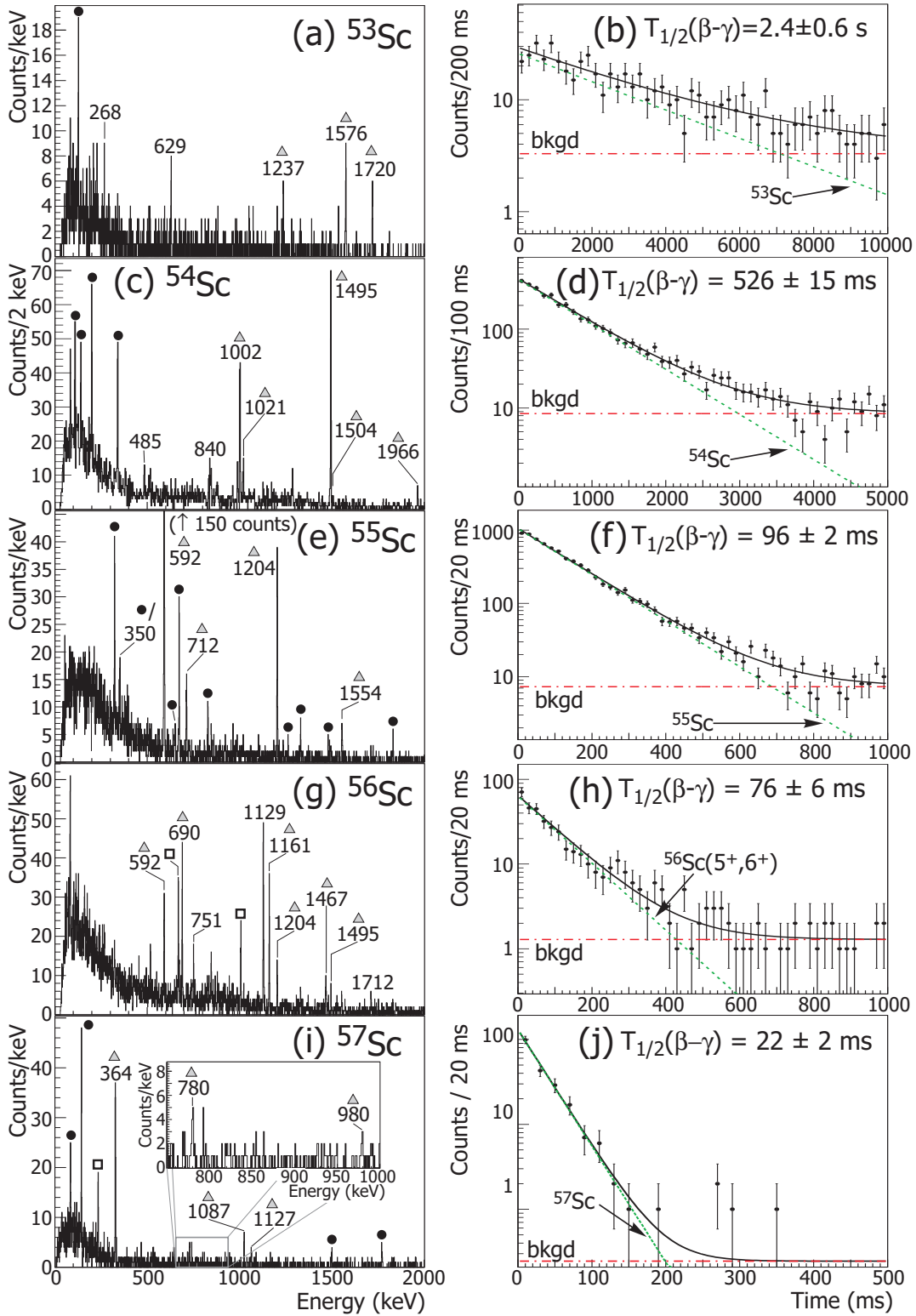}
\caption[Sc Beta-Delayed Gamma Spectra and Halflives]{(Color online) $\beta$-delayed 
$\gamma$-ray spectra (left column) and $\gamma$-gated half-life 
curves (right column) for (a, b) $^{53}$Sc, (c, d) $^{54}$Sc, (e, f) $^{55}$Sc, 
(g, h) $^{56}$Sc and (i, j) $^{57}$Sc.  The $\beta$-delayed $\gamma$-ray spectra 
cover transitions in the range 0-2~MeV.  The conditions under 
which individual spectra were obtained are discussed in the text.  
Observed transitions assigned to the decay of the Sc isotopes 
are marked by their energy in keV.  Transitions in the decay of 
the $\beta$ daughters are marked by a filled circle, while transitions 
in decays of the $\beta$ granddaughters are marked by a hollow square.  
In (c) and (g), the peak at 83~keV corresponds to a background 
transition in the $\beta$ decay of $^{56}$Cr.  The half-life curves made 
use of correlation times specified in the text, and 
required a coincident $\gamma$-ray (marked in the corresponding $\gamma$ 
spectrum by a shaded triangle).  Data in all cases were fitted with 
a single exponential decay and a constant 
background, where such a contribution was 
non-zero.}
\label{fig:ScAll}
\end{figure*}

Information regarding the $\beta$ decay of $^{53}$Sc is 
sparse.  Sorlin ${\it et~al.}$~\cite{Sorl98} reported the half-life 
of $^{53}$Sc as $>$ 3~s, and did not assign $\gamma$-ray transitions to 
the decay.  Given the anticipated long half-life, a correlation 
time of 10~s was used for analysis of this isotope, and only pixels 
on the edge of the DSSD were 
considered for the half-life and absolute $\gamma$-intensity determinations.  
On the other hand, the $\gamma$$\gamma$ coincidence analysis, 
which is less sensitive to 
random background, made use of data from the full DSSD surface.  
The $\beta$-delayed $\gamma$-ray spectrum for $^{53}$Sc of 
Fig.\ \ref{fig:ScAll}(a) contains six transitions, 
one corresponding to a known $\gamma$ ray 
in the decay of the daughter $^{53}$Ti~\cite{Park77}, and five 
assigned to the $\beta$ decay of 
$^{53}$Sc.  Four additional transitions, not apparent in the spectrum
of Fig.\ \ref{fig:ScAll}(a), were identified in the $\gamma$$\gamma$
coincidence analysis, and confirmed to be present in the $\gamma$-singles 
spectrum considering data over the full DSSD surface.  In the same 
$\gamma$-singles spectrum, two other weak transitions were identified, 
and assigned to the decay.  Transitions 
observed with energies of 292, 340, 629, 1237, and 1576~keV agree
with those observed in deep-inelastic work on $^{53}$Ti~\cite{Forn05}.

The decay curve for $\beta$-decay events correlated with 
$^{53}$Sc implantations, with the requirement of a 
high-energy decay $\gamma$ ray in 
coincidence with the decay event, is shown in 
Fig.\ \ref{fig:ScAll}(b).  The use of high-energy $\gamma$ 
rays above 1~MeV limited the background contribution.  
The fit of a single exponential 
decay to these data yielded a half-life of 2.4$\pm$0.6~s.

\begin{figure}[!tb]
\centering
\includegraphics[width = 0.48\textwidth]{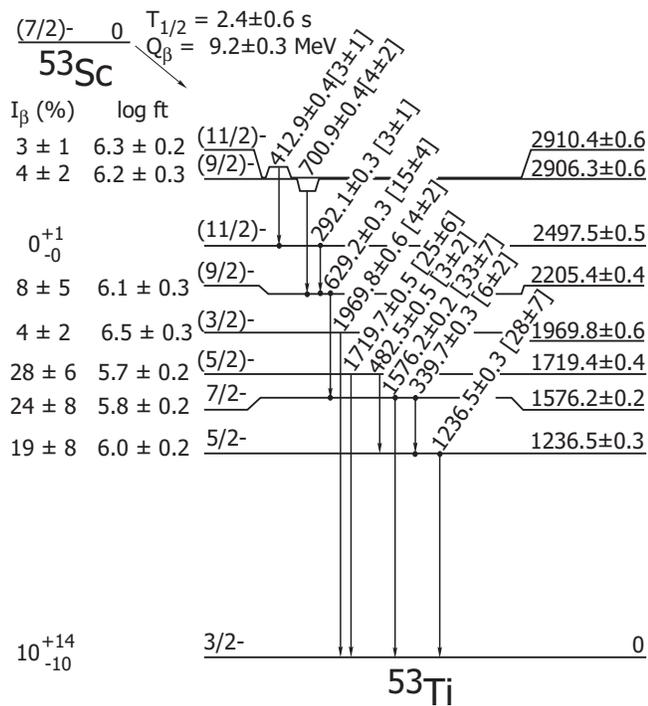}
\caption[53Sc Decay Scheme]{Proposed decay scheme for the $\beta$ 
decay of $^{53}$Sc to states in $^{53}$Ti.  The energy of each 
state is given in keV.  The number in brackets following the 
$\gamma$-ray energy is the absolute $\gamma$-ray intensity.  
The decay $Q$-value, ${\it Q_{\beta}}$, was deduced from data in 
Ref.~\cite{Audi03}. }
\label{fig:sc53DecayScheme}
\end{figure}

\begin{figure}[!tb]
\centering
\includegraphics[width = 0.48\textwidth]{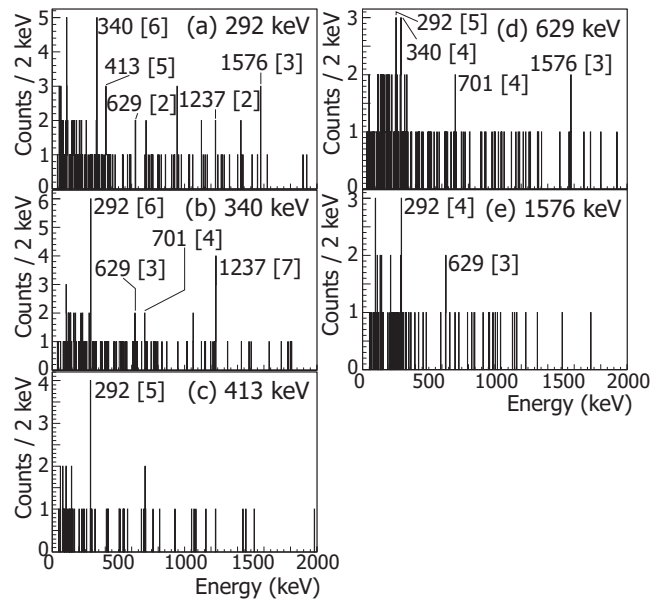}
\caption[53Sc Gamma Gamma]{$\gamma$$\gamma$ coincidence spectra gated on 
the (a) 292-keV, (b) 340-keV, (c) 413-keV, (d) 629-keV, and 
(e) 1576-keV transitions in coincidence with 
$^{53}$Sc-correlated $\beta$-decay events.  $\gamma$ rays are marked 
by their energy in keV, the number in square brackets following the 
energy is the number of counts in the peak.  These coincident 
spectra include data within a 10 s correlation time, for implants 
collected over the entire surface of the DSSD.}
\label{fig:sc53GammaGamma}
\end{figure}

The proposed decay scheme for levels populated in $^{53}$Ti by 
the decay of $^{53}$Sc is presented in Fig.\ \ref{fig:sc53DecayScheme}.  
The levels at 1237, 1576, 2205 and 2498~keV, and the associated
$\gamma$-ray transitions were positioned in accord with 
Ref.~\cite{Forn05} and confirmed by 
$\gamma$$\gamma$ coincidence data.  The spectrum of $\gamma$-ray 
transitions coincident with the 292-keV transition revealed 
a new $\gamma$ ray at
$\sim$413~keV [see Fig.\ \ref{fig:sc53GammaGamma}(a)], which was placed in 
cascade with the 292-keV line, depopulating a new state at 
2910~keV.  A 701-keV transition was observed in coincidence with 
the 629-keV line [Fig.\ \ref{fig:sc53GammaGamma}(d)], 
which was placed as connecting another new state at 2906~keV 
with the level at 2205~keV.  Both of these transitions, and the 
corresponding doublet of levels at $\sim$2910 keV have been confirmed by 
Cieplicka~${\it et~al.}$~\cite{Ciep09} using Gammasphere data 
from the reaction $^{48}$Ca + $^{238}$U, which populates states 
in $^{53}$Ti up to nearly 4 MeV.  The Gammasphere 
work also confirms the placement of the 
483-keV transition connecting the 1719- and 1237-keV levels, 
and is the source of the precise energy values for the 483- and 
1576-keV transitions.  
The absence of $\gamma$ rays coincident with the 
1720-keV and 1970-keV transitions lead to placement of these 
transitions feeding the ground state directly.  A 267.6$\pm$0.3-keV 
$\gamma$ ray (with an absolute intensity of 6$\pm$3\%) 
remains unplaced in the $^{53}$Ti level scheme.  

Spin and parity assignments proposed in Fig.\ 
\ref{fig:sc53DecayScheme} were adopted from the deep inelastic 
work of Fornal ${\it et~al.}$~\cite{Forn05}, and from 
comparison of the Gammasphere results with shell-model 
calculations~\cite{Ciep09}.  All assignments 
are consistent with the observed $\beta$ feeding pattern from the 
(7/2)$^{-}$ $^{53}$Sc ground state.  The apparent ground-state 
feeding is consistent with zero.

\subsection{\label{sec3:level2d}$^{54}$Sc $\beta$ decay}

The analysis of the $^{54}$Sc $\beta$ decay, with a half-life 
of order $\sim$400~ms~\cite{Lidd04}, considered a correlation 
time of 5~s, again restricted to include pixels 
towards the edge of the central implantation DSSD.  
The $\beta$-delayed $\gamma$-ray 
spectrum for $^{54}$Sc in the range 0 to 2~MeV is presented in 
Fig.\ \ref{fig:ScAll}(c).  Eleven transitions were 
identified in this spectrum, and the four lowest in energy were 
eliminated as candidates for transitions in the $^{54}$Sc decay, based on 
their apparent half-lives, and likely belong to the decay of the 
daughter nucleus, $^{54}$Ti.  $\gamma\gamma$ analysis indicates 
that these four transitions are in coincidence with one another, and 
the lowest energy transition at 108~keV corresponds to a known 
transition in $^{54}$V~\cite{Grzy98}.  The remaining seven $\gamma$ rays were 
assigned to the $\beta$ decay of $^{54}$Sc.  One additional transition 
was observed in the $\gamma$-spectrum considering data over the entire 
DSSD surface, and assigned to the decay of $^{54}$Sc.   The 
transitions observed at 1002, 1021 and 1495~keV were 
previously observed in $\beta$ decay~\cite{Lidd04} and 
deep-inelastic studies~\cite{Forn04}.  

The decay curve of $^{54}$Sc-correlated decay events with the 
additional requirement of a coincident 1002-, 1021-, 1495-, 1504- 
or 1966-keV $\gamma$-ray transition is given in 
Fig.\ \ref{fig:ScAll}(d).  A fit to these data including a single 
exponential decay and constant background yielded a half-life for 
$^{54}$Sc of 526$\pm$15~ms, longer than the value of 360$\pm$60~ms 
determined by Liddick ${\it et~al.}$~\cite{Lidd04}.  The 
discrepancy between the present value and previous half-life
measurements is attributed to the complex deconvolution of the 
$\beta$-decay curve.  A new half-life of 
2.1$\pm$1.0~s was determined here for the daughter $^{54}$Ti, 
which is slightly longer than the literature value~\cite{Dorf96}.  
Additionally, missing $\beta$ intensity is suggestive of a 
neutron-branching contribution of 16$\pm$9\%.  Inclusion of 
these factors in fitting a $\beta$-decay curve impacts the fit 
to data.  The additional $\gamma$-ray coincidence condition taken here 
permits a half-life determination that requires only two parameters, 
and extraction of a more accurate value.

\begin{figure}[!tb]
\centering
\includegraphics[width = 0.5\textwidth]{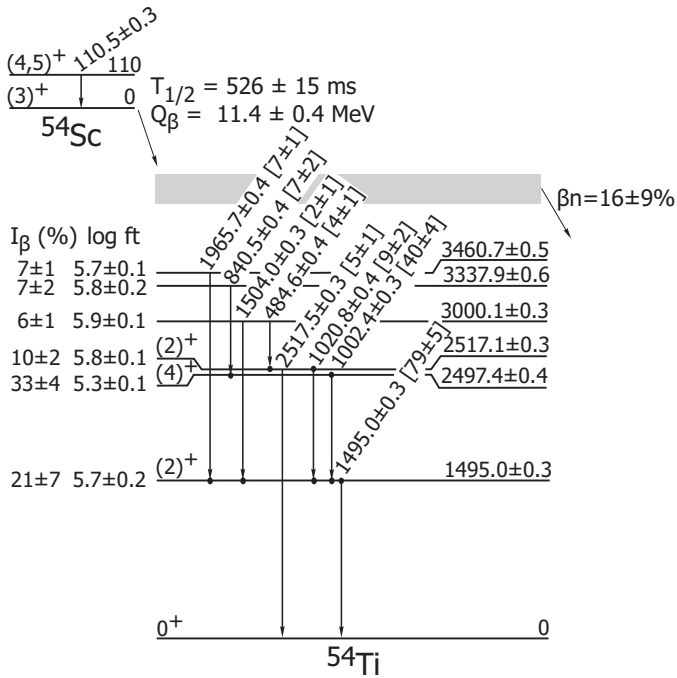}
\caption[54Sc Beta Decay Scheme]{Proposed decay scheme for 
$^{54}$Sc to states in $^{54}$Ti.  The energy of each state 
is given in keV.  The number in brackets following the 
$\gamma$-ray decay energy is the absolute $\gamma$-ray intensity.  
The ${\it Q_{\beta}}$ value was deduced from data in Ref.~\cite{Audi03}.}
\label{fig:sc54DecayScheme}
\end{figure}

The proposed decay scheme for $^{54}$Sc to levels in $^{54}$Ti 
is shown in Fig.\ \ref{fig:sc54DecayScheme}.  Placement 
of the 1002-, 1021- and 1495-keV transitions follows 
assignments made in previous works~\cite{Lidd04, Forn04}.  
The new transitions at 485, 840, 1504 and 1966~keV 
were placed based on $\gamma$$\gamma$ 
coincidences.  
Energy-sum relationships were used to place the 1504-keV 
cross-over transition connecting the states at 3000 and 
1495~keV, and the 2517-keV cross-over transition connecting the 
state at 2517 keV with the ground state.  

Apparent $\beta$ feedings and log $ft$ values deduced from the 
absolute $\gamma$ intensities are included in Fig.\ 
\ref{fig:sc54DecayScheme}.  The $J^{\pi}$ of the $^{54}$Sc ground 
state was previously limited to (3,4)$^{+}$~\cite{Lidd04}.   Apparent 
$\beta$ feeding is observed to both the 2$_{1}^{+}$ and 4$_{1}^{+}$ states 
previously identified in Ref.~\cite{Forn04}, at 1495 and 2497~keV 
respectively.  The suggested allowed decay to both of these states further restricts the spin 
and parity for the $^{54}$Sc ground state to 3$^{+}$.  No feeding is 
expected to the 0$^{+}$ $^{54}$Ti ground state according to the selection 
rules for allowed $\beta$ decay, assuming a $J^{\pi}$ = 3$^{+}$ 
ground state for $^{54}$Sc.  $J^{\pi}$ values 
for the states with energy 1495, 2497 and 2517 keV were adopted from previous 
works~\cite{Lidd04, Forn04}.  Apparent allowed $\beta$ feeding from 
the (3)$^{+}$ $^{54}$Sc ground state limits the spin and parity of 
the newly placed levels at 3000, 3338 and 3461~keV to $J^{\pi}$ = 
2$^{+}$, 3$^{+}$, or 4$^{+}$. 

\subsection{\label{sec3:level2e}$^{54}$Sc isomeric decay}

\begin{figure}[!tb]
\centering
\includegraphics[width = 0.4\textwidth]{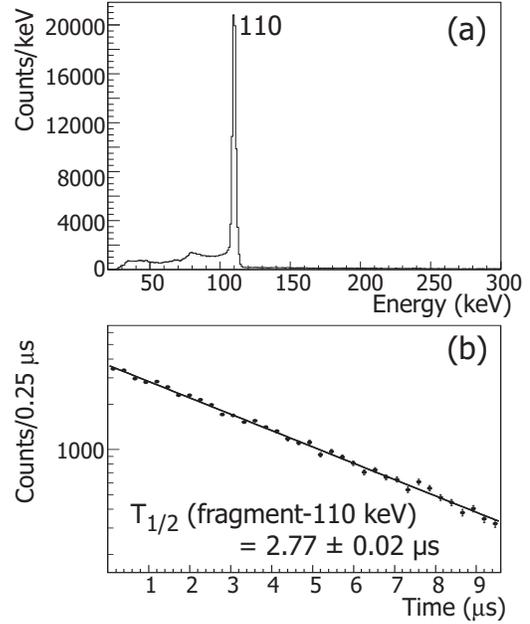}
\caption[54Sc Isomer Summary]{(a) $\gamma$-ray spectrum collected within 
a 20-$\mu$s window following a $^{54}$Sc implant.  The 110-keV isomeric 
transition observed here was previously reported in Ref.~\cite{Grzy98}, 
and confirmed in Ref.~\cite{Lidd04}.  (b) Time dependence of the 
110-keV transition.  The decay curve was fitted with a single 
exponential decay to yield a half-life of 2.77$\pm$0.02~$\mu$s.}
\label{fig:sc54Isomer}
\end{figure}

The isomeric $\gamma$-ray spectrum collected within 20~$\mu$s 
of a $^{54}$Sc implantation event is presented in Fig.\ 
\ref{fig:sc54Isomer}(a).  The 110-keV isomeric 
transition was initially reported by Grzywacz 
${\it et~al.}$~\cite{Grzy98}, and 
assigned E2 multipolarity.  The transition was observed here with 
high statistics, and an 
improved half-life of 2.77$\pm$0.02~$\mu$s was deduced [Fig.\ 
\ref{fig:sc54Isomer}(b)].  This more precise value is consistent with the 
result of Grzywacz ${\it et~al.}$, and an the E2 multipolarity 
assignment based on comparison to Weisskopf lifetime estimates.  
Based on the (3)$^{+}$ spin and parity of the $^{54}$Sc ground state 
and an E2 isomeric transition, the limits on $J^{\pi}$ for the 110-keV 
level are $J^{\pi}$ = 1$^{+}$ or 5$^{+}$.  $J$ = 1 is excluded as 
this state was not directly populated following the $\beta$ decay of 
$^{54}$Ca.  Thus, under the assumption of an E2 multipolarity for the 
transition, the 110-keV isomeric state in $^{54}$Sc is tentatively 
assigned a spin and parity of 5$^{+}$, as originally suggested by 
Grzywacz ${\it et~al.}$~\cite{Grzy98}.  However, this assignment assumes 
a single-particle nature for the final and initial states.  More complicated wavefunctions, involving configuration mixing, open the window for a 4$^{+}$ spin and parity assignment for the 110-keV isomeric state.  This possibility is explored further in Section IV.B. 

\subsection{\label{sec3:level2f}$^{55}$Sc $\beta$ decay}

A correlation time of 1~s was used for the analysis of events 
associated with $^{55}$Sc, based on previous half-life 
measurements~\cite{Sorl98, Lidd04, Mant08} indicative of a 
$\sim$100~ms half-life.  Analysis of the $\beta$-gated 
$\gamma$-ray spectrum was again restricted to pixels towards 
the edge of the central implantation DSSD.  
The spectrum in the range of 0 to 2~MeV is given in 
Fig.\ \ref{fig:ScAll}(e).  A total of 13 $\gamma$-ray 
transitions were identified in this spectrum; of these, 9 
correspond to known $\gamma$ rays in the decay of the daughter isotope, 
$^{55}$Ti.  The remaining 4 transitions have been 
assigned to the $\beta$ decay of 
$^{55}$Sc.  One peak in the $\gamma$-singles spectrum, at 
349.6$\pm$0.7~keV, is very broad, and is likely a doublet.  
A known transition in the $\beta$ decay of 
$^{55}$Ti, with energy 349.3$\pm$0.6~keV, may contribute to this peak.  
The transition with energy 592~keV corresponds 
to the $\gamma$ ray previously observed in the $\beta$ decay of $^{55}$Sc by 
Liddick ${\it et~al.}$~\cite{Lidd04}.  
States in $^{55}$Ti have been studied by deep inelastic scattering with 
Gammasphere at the ATLAS facility~\cite{Zhu07}.  The $\gamma$ rays 
reported here with energies 350, 592, 1204 and 1554~keV agree with transitions 
observed in that work.  

A summed decay curve from gates on the $\gamma$ rays in $^{55}$Ti with energies 
592, 712, 1204 and 1554~keV is given in Fig.\ \ref{fig:ScAll}(f).  
This curve was fitted with a single 
exponential decay and constant background, and the resulting half-life 
of 96$\pm$2~ms is consistent with previous 
measurements~\cite{Lidd04, Mant08}.

\begin{figure}[!tb]
\centering
\includegraphics[width = 0.48\textwidth]{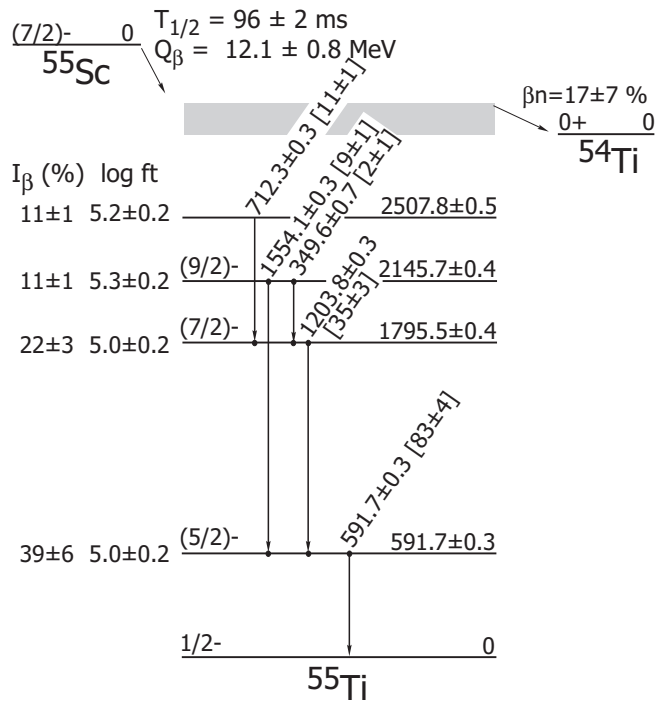}
\caption[55Sc Decay Scheme]{Decay scheme for $^{55}$Sc to states 
in $^{55}$Ti.}
\label{fig:sc55DecayScheme}
\end{figure}

The level scheme for $^{55}$Ti populated following $\beta$ decay of 
$^{55}$Sc is illustrated in Fig.\ \ref{fig:sc55DecayScheme}.  Placement of 
the 350-, 592-, 1204- and 1554-keV transitions is based on previous 
deep inelastic results~\cite{Zhu07}, which established the yrast 
level structure for $^{55}$Ti up to $J^{\pi}$=(19/2)$^{-}$.  One 
additional transition, not seen in Ref.~\cite{Zhu07}, with energy 712~keV 
was observed in coincidence with the 1204- and 592-keV transitions.  
This 712-keV transition was placed depopulating a new state 
with energy 2508~keV.  

Tentative $J^{\pi}$ values 
were assigned for the ground state and excited states at 592, 1796 
and 2146~keV in Ref.~\cite{Zhu07}, based on comparisons to 
shell model calculations with the 
GXPF1A interaction~\cite{Honm05}.  The assignments proposed in that work 
included $J^{\pi}$ = 1/2$^{-}$ for the $^{55}$Ti ground state, which 
has recently been confirmed by a one-neutron knockout measurement 
from $^{56}$Ti to states in $^{55}$Ti performed at GSI~\cite{Maie09}.  
Such a $J^{\pi}$ assignment would exclude allowed $\beta$ decay from 
the $^{55}$Sc ground state populating the $^{55}$Ti ground state directly.  
Thus, the missing $\beta$ intensity of 17$\pm$7\% suggests the possible presence 
of a $\beta$n branch in the $^{55}$Sc decay.  Absence of transitions in $^{54}$Ti 
(see Fig.\ \ref{fig:sc54DecayScheme}) in the $\beta$-delayed $\gamma$-ray
spectrum may indicate that the $\beta$n branch mainly populates the 
$^{54}$Ti ground state 
directly.  The spin and parity assignments proposed by Zhu 
${\it et~al.}$~\cite{Zhu07} are adopted here.  Apparent allowed 
$\beta$ decay to the newly placed state at 2508~keV limits this 
state to $J^{\pi}$ = 5/2$^{-}$, 7/2$^{-}$ or 9/2$^{-}$.

\subsection{\label{sec3:level2g}$^{56}$Sc $\beta$ decay}

Analysis of the $\beta$ decay of $^{56}$Sc was completed using a 
correlation time of 1 s, and the full DSSD surface was 
considered.  The $\beta$-delayed $\gamma$-ray 
spectrum for $^{56}$Sc in the range of 0 to 2~MeV can be found in 
Fig.\ \ref{fig:ScAll}(g), where nine transitions 
have been assigned to the decay of $^{56}$Sc, and are summarized in Table 
\ref{tab:sc56Gammas}.  Two additional transitions were assigned to 
the decay of the granddaughter nuclide, $^{56}$V.  The transitions 
with energies 592, 690, 751, 1129 and 1161~keV were previously observed 
in both $\beta$ decay~\cite{Lidd04} and deep-inelastic 
experiments~\cite{Forn04}.  It was noted in Ref.~\cite{Lidd04} that 
the transition with energy 592.3$\pm$0.5~keV was within the error of a known 
transition in $^{55}$Ti, and the possibility of $\beta$-delayed neutron 
emission was suggested.  Another 
known transition in $^{55}$Ti, at 1204~keV~\cite{Zhu07} is also present in the 
delayed $\gamma$-ray spectrum for $^{56}$Sc, confirming 
$\beta$-delayed neutron emission in this decay.  

\begin{table}[!tb]
\caption{Energies and relative intensities of $\beta$-delayed 
$\gamma$ rays assigned to the decay of $^{56}$Sc. Observed half-lives for 
$\gamma$-gated decay curves are also included.}
\label{tab:sc56Gammas}
\begin{ruledtabular}
\begin{tabular}{cccccc}
$E_{\gamma}$ & $I^{absolute}_{\gamma}$ & Decay & Initial & Final & $T_{1/2}$ \\
(keV) & (\%) & Mode & State & State & (ms) \\
 & & & (keV) & (keV) & \\
\hline
$591.7\pm0.3$ & $14\pm2$ & $\beta$n & 592 & 0 & $78\pm25$ \\
$689.6\pm0.3$ & $18\pm2$ & $\beta$ & 2979 & 2289 & $73\pm10$ \\
$750.9\pm0.4$ & $8\pm2$ & $\beta$ & 1880 & 1129 & $24\pm7$ \\
$1128.7\pm0.3$ & $48\pm4$ & $\beta$ & 1129 & 0 & $51\pm6$ \\
$1160.6\pm0.3$ & $30\pm3$ & $\beta$ & 2289 & 1129 & $78\pm9$\\
$1203.5\pm0.3$ & $8\pm1$ & $\beta$n & 1796 & 592 & $68\pm19$ \\
$1466.8\pm0.3$ & $6\pm1$ & $\beta$ & - & - & $60\pm13$ \\
$1494.8\pm0.3$ & $3\pm1$ & $\beta$ & 4474 & 2979 & $150\pm44$ \\
$1711.6\pm0.3$ & $3\pm1$ & $\beta$ & - & - & $29\pm10$ \\   
\end{tabular}
\end{ruledtabular}  
\end{table}  

Liddick ${\it et~al.}$~\cite{Lidd04} proposed two $\beta$-decaying 
states in $^{56}$Sc: a low-spin state with $T_{1/2}$ = 35$\pm$5~ms, and 
a higher-spin level with slightly longer $T_{1/2}$ = 60$\pm$7 ms.  
Given the different spin values, the two $\beta$-decaying states 
were observed to populate different levels in the 
$^{56}$Ti daughter.  Gated decay curves 
were generated for each $\gamma$-ray transition identified in Fig.\ 
\ref{fig:ScAll}(g), and fitted with a single exponential decay and constant 
background.  The resultant half-life values are summarized in Table 
\ref{tab:sc56Gammas}.  Values deduced from the decay curves gated on the 690- 
and 1161-keV $\gamma$ rays, which depopulate levels at 2979 and 
2289~keV respectively, are in agreement with the previous 
half-life determination for the higher-spin 
isomer~\cite{Lidd04}.  The transitions assigned to the 
$\beta$n decay also have half-lives consistent with decay 
from the higher-spin state, as do the transitions with energies 
1467 and 1495~keV.  The weighted average of the half-lives for 
transitions populated solely by the higher-spin isomer is 75$\pm$6~ms.

The half-life of the lower-spin, $\beta$-decaying state~\cite{Lidd04} 
was previously deduced from a two-component fit to the 
decay curve gated on the 1129-keV $\gamma$ 
transition, which depopulates a state fed by both $\beta$-decaying 
states in $^{56}$Sc.  Two 
newly-identified $\gamma$-ray transitions at 751 and 1712 keV decay with 
half-lives similar to that extracted for the lower-spin isomer.  
These $\gamma$ rays apparently depopulate states 
fed exclusively by the lower-spin isomer.  The weighted average of the 
half-lives for the 751- and 1712-keV transitions yields a half-life 
for the lower-spin isomer of 26$\pm$6~ms.

\begin{figure*}[!tb]
\centering
\includegraphics[width = 1\textwidth]{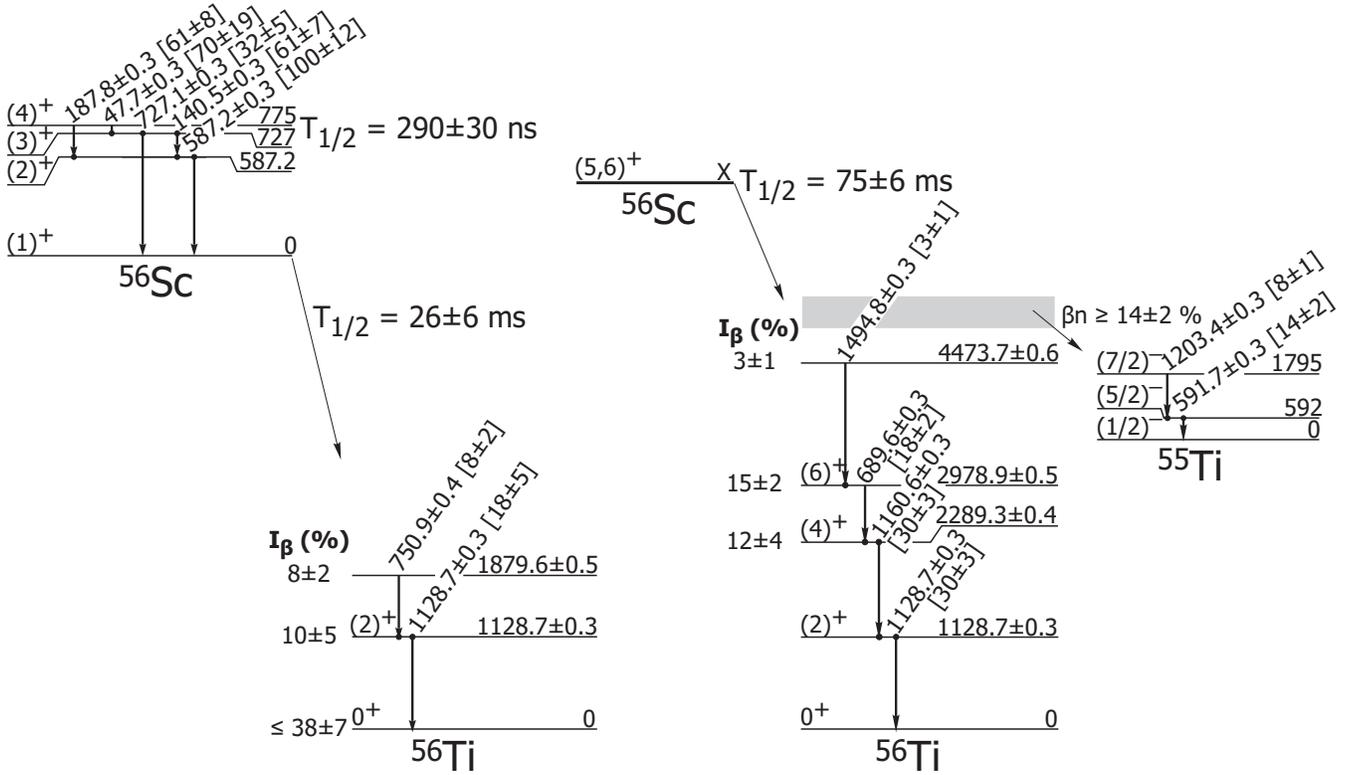}
\caption[56Sc Isomer Decay Scheme]{Proposed level scheme for $^{56}$Sc, 
following the $\gamma$ decay of the isomeric transition at 775~keV, and 
levels in $^{55,56}$Ti populated following $\beta$-decay of $^{56}$Sc.  
The number in brackets following the isomeric $\gamma$-ray energies are the relative 
$\gamma$ intensities.  The numbers following the $\beta$-delayed $\gamma$-ray energies are the absolute $\gamma$-ray intensities.  The relative energies of the two $\beta$-decaying states is not known.  The energy of the higher-spin $\beta$-decaying state, denoted by ``X'' may be negative.  $Q_{\beta}$ for this decay is 13.7$\pm$0.7~MeV~\cite{Audi03}.}
\label{fig:sc56DecayScheme}
\end{figure*}

Levels in $^{56}$Ti and $^{55}$Ti populated in the decay of the two 
$\beta$-decaying states in $^{56}$Sc are shown in Fig.\ 
\ref{fig:sc56DecayScheme}.  States in $^{55}$Ti are known from the 
present $^{55}$Sc $\beta$-decay results, previous $\beta$-decay 
studies~\cite{Lidd04}, and deep-inelastic work~\cite{Zhu07}.  
The three states at 1129, 2289 and 2979~keV in $^{56}$Ti were 
previously identified and assigned tentative spin and parity 
values~\cite{Lidd04, Forn04}, which were adopted 
in Fig.\ \ref{fig:sc56DecayScheme}.  
Observed $\gamma$$\gamma$ coincidences between the 1129- 
and 751-keV transitions suggest the placement 
of an additional state populated by the lower-spin 
$\beta$-decaying state at 1880~keV.  In addition to 
populating the two states at 2289 and 2979~keV directly, 
observed $\gamma$$\gamma$ coincidences between the 1495- 
and 690-keV transitions suggest that the higher-spin $\beta$-decaying 
state populates a level at 4474~keV.  Only the 1467- and 1712-keV 
$\gamma$ transitions remain unplaced in the present level scheme for $^{56}$Ti.  

Absolute $\gamma$-ray intensities were determined for the $\beta$-delayed 
$\gamma$-ray transitions in the $^{56}$Sc decay by comparison of the number 
of observed $\gamma$ rays, adjusted for the absolute efficiency of SeGA, with 
the number of observed $^{56}$Sc decays.  The number of parent decays in 
this case was determined using the total number of implantations, taken from 
the particle identification, and the average $\beta$-detection efficiency of 
11.4$\pm$0.4\%.  The absolute $\gamma$-ray intensities are included in 
Table~\ref{tab:sc56Gammas}, and in the decay scheme of 
Fig.~\ref{fig:sc56DecayScheme}.  Apparent $\beta$ branches were deduced 
from the absolute $\gamma$-ray intensities.  Decay from the higher-spin 
isomer apparently populates both the 4$^{+}$ and 6$^{+}$ states in $^{56}$Ti.  
This suggests that the higher-spin isomer has $J^{\pi}$=5$^{+}$, in 
contradiction to the previous assignment of (6,7)$^{+}$ by Liddick 
\textit{et~al.}~\cite{Lidd04}.  However, the large $Q_{\beta}$ value for the 
decay allows the (4)$^{+}$ state at 2289~keV to be populated by cascades from 
above, resulting in the observed intensity difference between the 690- and 
1161-keV transitions.  Given this possibility, the higher-spin $\beta$-decaying 
state has been tentatively assigned as (5,6)$^{+}$.  The state with an 
energy of 4474~keV, also populated by the higher-spin $\beta$-decaying 
state, has $J^{\pi}$ limited to 4$^{+}$, 5$^{+}$, 6$^{+}$ or 7$^{+}$, depending 
on the spin and parity of the higher-spin $\beta$-decaying state in $^{56}$Sc.

Decay from the lower-spin $\beta$-decaying state in $^{56}$Sc apparently 
directly feeds the first excited 2$^{+}$ state in $^{56}$Ti, which limits 
the spin and parity of this $\beta$-decaying state to 1$^{+}$, 2$^{+}$ 
or 3$^{+}$.  The $J^{\pi}$ for the $^{56}$Sc ground state can be further 
restricted if there is direct $\beta$ feeding to the $^{56}$Ti ground 
state.  The absolute intensity of the 592-keV transition in the $\beta$n 
daughter $^{55}$Ti suggests a lower-limit for the neutron branching 
ratio of 14$\pm$2\%.  Even under the assumption that the two unplaced 
$\gamma$ transitions directly populate the $^{56}$Ti ground state, 
there is 29$\pm$7\% of the $\beta$ intensity unaccounted for.  
Two scenarios can account for the missing intensity:  either direct 
feeding to the $^{56}$Ti ground state or $\beta$n decay populating 
the ground state of $^{55}$Ti directly.  These possibilities were 
investigated by reanalyzing the $\beta$ decay of $^{56}$Sc with the 
longer correlation time of 5~s.  With a longer correlation time, it 
was possible to compare the intensities of $\gamma$-ray transitions 
in the decay of the $\beta$n daughter $^{55}$Ti (672.5~keV, 
$I_{\gamma}^{absolute}=44\pm4$\%~\cite{Mant03}), and the $\beta\beta$ granddaughter 
$^{56}$V (668.4~keV, $I_{\gamma}^{absolute}=26\pm2$\%~\cite{Mant03_2}).  The ratio of $
\beta$ to $\beta$n decays in $^{56}$Sc can be determined by the ratio 
of the intensities of the 668- and 673-keV transitions, which are detected 
with nearly the same efficiency in SeGA.  Approximately 70\% of the decays 
of $^{56}$Sc populate states in $^{56}$Ti.  Thus, the missing $\beta$ intensity 
cannot be fully accounted for by $\beta$n decay to the ground state of $^{55}$Ti, 
and there is apparent direct population of the 0$^{+}$ $^{56}$Ti ground state, which 
must be from the lower-spin $\beta$-decaying state.  This suggests a spin and parity 
for the lower-spin $\beta$-decaying state of $J^{\pi}$=1$^{+}$.  Direct 
feeding from the (1)$^{+}$ $\beta$-decaying state limits $J^{\pi}$ of the state 
at 1880~keV to (0,1,2)$^{+}$.

\subsection{\label{sec3:level2h}$^{56}$Sc isomer decay}

\begin{figure}[!tb]
\centering
\includegraphics[width = 0.48\textwidth]{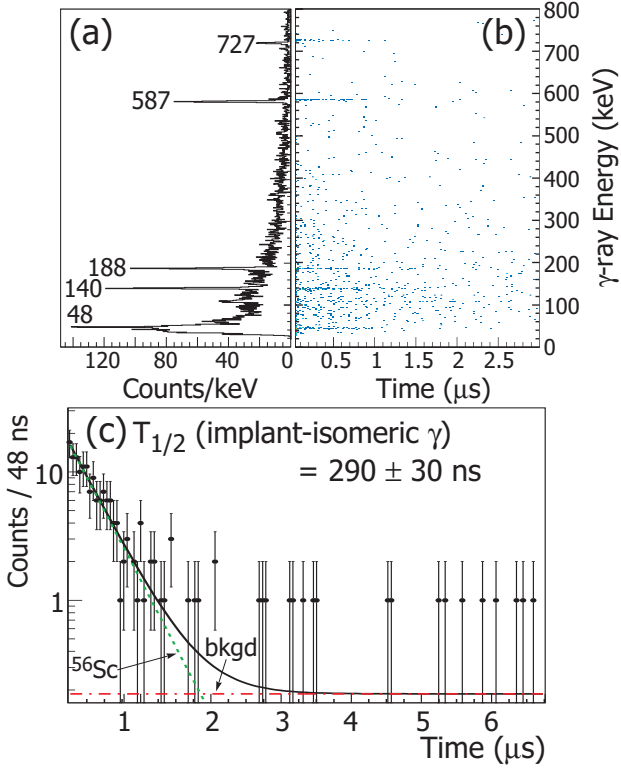}
\caption[56Sc Isomer Gammas and Decay]{(Color online) (a) The prompt 
$\gamma$-ray spectrum collected within the first 20 $\mu$s following a 
$^{56}$Sc implant.  The $\gamma$ rays assigned to the isomer decay in 
$^{56}$Sc are marked by their energies in keV.  The additional peak 
visible in the spectrum at 110 keV is contamination from the isomer decay 
of $^{54}$Sc.  (b) A plot of $\gamma$-ray energy as a function of time 
elapsed after the corresponding $^{56}$Sc implantation event.  The five 
isomer $\gamma$ lines are visible, decaying with comparable lifetimes.  
(c) A projection of (b) onto the time axis, gated on the five 
identified $\gamma$-ray transitions.  The resultant decay curve was 
fitted with a single exponential decay and constant background.}
\label{fig:sc56IsomerGammas}
\end{figure}

The isomeric $\gamma$-ray spectrum collected within a 20-$\mu$s time 
window following a $^{56}$Sc implantation is presented in Fig.\ 
\ref{fig:sc56IsomerGammas}(a).  Five transitions were observed.  
Those with energies 140, 188 and 587 keV were previously 
reported by Liddick ${\it et~al.}$~\cite{Lidd04}.  

The observed $\gamma$-ray energy as a function of the time elapsed 
between a $^{56}$Sc implantation event and prompt $\gamma$ emission 
is presented in Fig.\ \ref{fig:sc56IsomerGammas}(b).  The five 
transitions appear to decay with the same half-life.  The projection of Fig.\ 
\ref{fig:sc56IsomerGammas}(b) onto the time axis, gated on the five 
isomeric transitions, is included in Fig.\ \ref{fig:sc56IsomerGammas}(c).  
The resultant decay curve was fitted with a 
single exponential decay and a constant background, resulting in a 
half-life of 290$\pm$30~ns.

\begin{figure}[!tb]
\centering
\includegraphics[width = 0.5\textwidth]{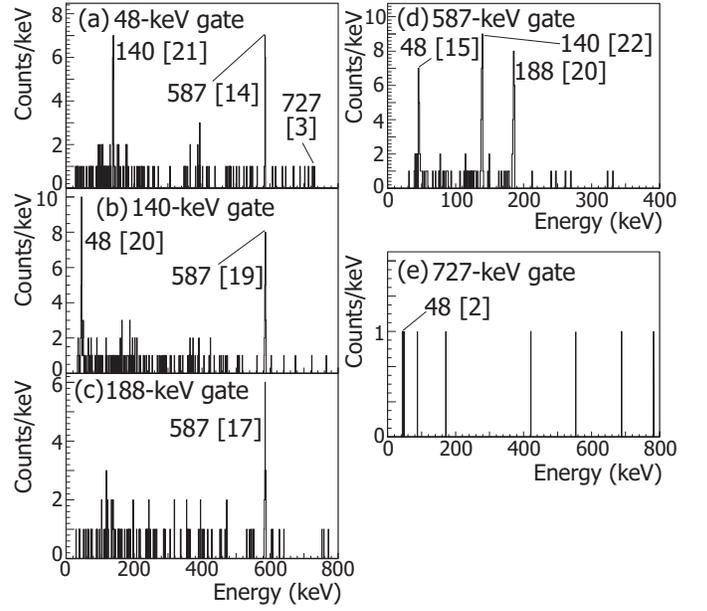}
\caption[56Sc Isomer Gamma Coincidences]{$\gamma$$\gamma$ coincidence 
gates for the (a) 48-keV, (b) 140-keV, (c) 188-keV, (d) 587-keV and 
(e) 727-keV transitions in the 20-$\mu$s time window following a 
$^{56}$Sc implant.  $\gamma$ rays are marked by their energy in keV 
and the number in square brackets is the number of counts in the peak.}
\label{fig:sc56IsomerGammaGamma}
\end{figure}

The low-energy structure of $^{56}$Sc, populated by isomeric 
decay (as presented in Fig. \ref{fig:sc56DecayScheme}), was based 
on observed $\gamma$$\gamma$ coincidences (see 
Fig.\ \ref{fig:sc56IsomerGammaGamma}), as well as relative intensity and 
energy-sum relationships.

\begin{figure}[!tb]
\centering
\includegraphics[width = 0.4\textwidth]{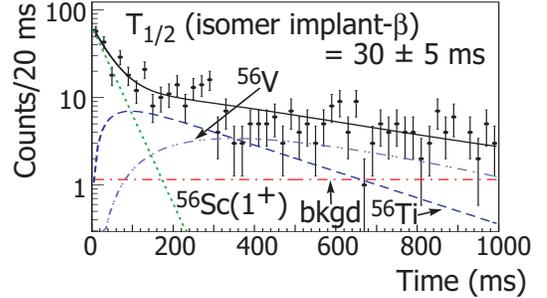}
\caption[56Sc Decay Curves]{(Color online) Decay curve for 
$^{56}$Sc-correlated $\beta$ decays considering only $^{56}$Sc 
implants coincident with identified isomeric $\gamma$-ray transitions.  
The decay curve is fitted with a single exponential decay 
and a constant background, 
and includes contributions from the 
growth and decay of the $\beta$-decay daughter 
($^{56}$Ti, $T_{1/2} = 200\pm5$~ms~\cite{Mant03}) and 
granddaughter ($^{56}$V, $T_{1/2} = 216\pm4$~ms~\cite{Mant03_2}).}
\label{fig:sc56DecayCurves}
\end{figure}

A novel approach was taken to determine which of the two 
$\beta$-decaying states in $^{56}$Sc is populated in the isomeric decay from the 
775-keV level.  A half-life curve was constructed considering only $\beta$-decay 
events correlated with $^{56}$Sc implantation events that were 
in turn correlated with one of the five prompt $\gamma$ rays.  The resultant 
decay curve is presented in Fig.\ \ref{fig:sc56DecayCurves}.  The 
deduced short half-life of 30$\pm$5~ms provides evidence that the 
isomeric transitions populate the lower-spin $\beta$-decaying state.  

The tentative spin and parity assignment of the lower-spin $\beta$-decay isomer in 
$^{56}$Sc as 1$^{+}$ permits tentative $J^{\pi}$ assignments to the remaining 
levels populated in decay of the isomeric state 
in $^{56}$Sc.  The 188-keV transition, 
which directly depopulates the isomeric level, is proposed to be of E2 
character, based on the comparison to Weisskopf half-life estimates, 
and the competing 48-keV transition to be of the M1 type.  It follows that 
the 140-keV transition must also have M1 multipolarity.  The observed 
intensities of the 727-keV and 140-keV transitions then suggest 
the former $\gamma$ ray to be an E2 transition.  
Finally, the remaining 
587-keV transition must have M1 multipolarity.  These assignments lead to 
the tentatively assigned $J^{\pi}$ = 2$^{+}$, 3$^{+}$ and 4$^{+}$ values 
for the states in $^{56}$Sc with energies 587, 727 and 775~keV above the 
lower-spin $\beta$-decaying state respectively, again assuming that this 
state in $^{56}$Sc has $J^{\pi}$=1$^{+}$ quantum numbers.

\subsection{\label{sec3:level2i}$^{57}$Sc $\beta$ decay}

$^{57}$Sc was studied in a separate experiment from that used to deduce the 
results for the isotopes previously discussed.  However, 
the decay of $^{57}$Sc to states in 
$^{57}$Ti was analyzed in a similar manner.  A correlation time of 
500~ms was used for the analysis of this decay, which had a previous 
half-life measurement of order tens of milliseconds.  Pixels over 
the entire DSSD surface were considered in the analysis.  

The $\beta$-delayed $\gamma$-ray spectrum for the decay of $^{57}$Sc 
is presented in Fig.\ \ref{fig:ScAll}(i).  
Four transitions in this spectrum were attributed to the $\beta$ decay 
of the $^{57}$Ti daughter nucleus, and one transition was known from 
the $\beta$ decay of the $^{57}$V granddaughter.  The remaining 
five $\gamma$ rays in Fig.\ \ref{fig:ScAll}(i) were attributed 
to the $\beta$ decay of $^{57}$Sc, and are listed in Table 
\ref{tab:Sc57Gammas}.  No previous information on $\beta$-delayed 
$\gamma$ rays in $^{57}$Sc is available from the literature.  
The transition at 1127~keV is close 
in energy to a known transition in $^{56}$Ti, at 
1128.7$\pm$0.3~keV~\cite{Lidd04, Forn04}, and may suggest a 
$\beta$-delayed neutron branch from $^{57}$Sc to levels in $^{56}$Ti.  

\begin{table}[!tb]
\caption{Energies and absolute intensities of $\beta$-delayed $\gamma$ rays 
assigned to the decay of $^{57}$Sc.}
\label{tab:Sc57Gammas}
\begin{ruledtabular}
\begin{tabular}{cc}
$E_{\gamma} (keV)$ & $I^{abs}_{\gamma}$ (\%) \\
\hline
$364.2 \pm 0.4$ & $76 \pm 8$ \\
$780.4 \pm 0.5$ & $9 \pm 3$ \\
$979.6 \pm 0.5$ & $5 \pm 1$ \\
$1087.1 \pm 0.4$ & $21 \pm 3$ \\
$1127.1 \pm 0.5$ & $12 \pm 2$ \\
\end{tabular}
\end{ruledtabular}  
\end{table}

The decay curve for $^{57}$Sc-correlated $\beta$ decays 
with the additional requirement of a decay $\gamma$-ray coincidence 
is given in Fig.\ \ref{fig:ScAll}(j).  The decay curve was 
fitted with a single exponential decay and a constant background, 
and the deduced half-life was 22$\pm$2~ms.  The new half-life value 
is slightly longer than the previous $^{57}$Sc half-life measurement of 
13$\pm$4~ms~\cite{Sorl03}, which did not include a $\gamma$-ray tag.

No $\gamma$$\gamma$ coincidence data were available due to low 
statistics.  The absolute intensity for the 364-keV $\gamma$
ray suggests that it directly populates the $^{57}$Ti ground state, 
and it has been placed as a ground-state transition.  No 
other conclusions regarding the low energy structure of $^{57}$Ti
could be drawn from the present data.

%%%%%%%%%%%%%%%%%%%%%%  DISCUSSION   %%%%%%%%%%%%%%%%%%%%%%%%%%%%%%%%%%%%

\section{\label{sec4:level1}Discussion}

The low-energy levels of the $_{21}$Sc isotopes can be 
discussed, in the context of the extreme single-particle 
model, as a coupling of the odd $1f_{7/2}$ proton 
to the states in the $_{20}$Ca core isotopes.  The success 
or failure of simple coupling 
schemes in explaining the low-energy structure of the Sc isotopes 
can also provide indirect indications regarding the presence or absence of 
subshell closures in the Ca core nuclei.  The structure of the Sc 
isotopes is first discussed in the framework of this coupling, and 
then compared to the results of shell model calculations.  Results 
in the Ti isotopes are discussed in terms of the systematic evolution 
of structure along the $Z$=22 isotopic chain.   

\subsection{\label{sec4:level2a}Low-energy structure of $^{53}$Sc}

\begin{figure}[!tb]
\centering
\includegraphics[width = 0.4\textwidth]{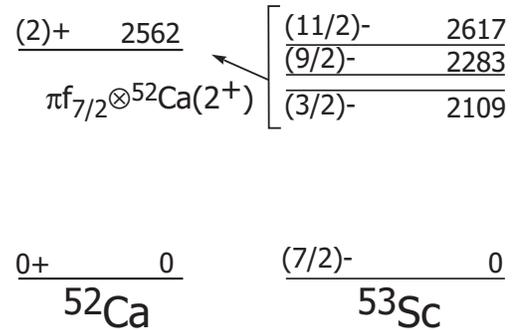}
\caption[53Sc coupling scheme]{The comparison of the known levels in 
$^{53}$Sc~\cite{Bhat09, Craw09} 
to states in $^{52}$Ca suggests that $^{53}$Sc is accurately described by 
weak coupling of the odd $\pi$$1f_{7/2}$ proton to the $^{52}$Ca core.}
\label{fig:sc53_coupling}
\end{figure}

The low-energy structure of $^{53}$Sc 
can be interpreted as a weak coupling of the valence 
$1f_{7/2}$ proton to states in $^{52}$Ca, which, assuming a robust 
$N$=32 subshell closure, can be viewed as a doubly magic 
core~\cite{Gade06, Craw09}.  Coupling of the valence proton to the 
first 2$^{+}$ state in $^{52}$Ca should produce a quintet of states in $^{53}$Sc with 
$J^{\pi}$ values ranging from 3/2$^{-}$ to 11/2$^{-}$, at energies 
centered around $\sim$2.6~MeV.  Studies of $^{53}$Sc via deep inelastic 
reactions~\cite{Bhat09} have populated 
9/2$^{-}$ and 11/2$^{-}$ states near this energy, which are likely members 
of the $\pi$$1f_{7/2}$$\otimes$2$^{+}$ multiplet.  The $\beta$ decay of 
$^{53}$Ca was observed to populate only one state in $^{53}$Sc.  The population of a single 
state in the decay of $^{53}$Ca decay limits the spin and parity of the 
$^{53}$Ca ground state to $J^{\pi}$=1/2$^{-}$, as allowed $\beta$ decay from a 
$J^{\pi}$ = 3/2$^{-}$ or 5/2$^{-}$ $^{53}$Ca 
ground state would be expected to populate several levels of the 
$\pi$$1f_{7/2}$$\otimes$2$^{+}$ multiplet.  The single new level at 2109 keV 
populated in the decay from a $J^{\pi}$=1/2$^{-}$ $^{53}$Ca ground state is 
a likely candidate for the 3/2$^{-}$ member 
of the $\pi$$1f_{7/2}$$\otimes$2$^{+}$ multiplet.  
The three known levels in $^{53}$Sc (see Fig.\ \ref{fig:sc53_coupling}) reside 
at just over 2~MeV, nearly the energy expected within 
the weak coupling framework.  The success of this scheme in 
describing the structure of $^{53}$Sc provides further support for the 
robust nature of the $N$=32 subshell closure in the $_{20}$Ca isotopes.  

\subsection{\label{sec4:level2b}Low-energy structure of even-A Sc isotopes}

The structure of the even-$A$ $_{21}$Sc isotopes can be described in 
terms of coupling of the valence $1f_{7/2}$ proton particle 
with $2p_{3/2}$, $2p_{1/2}$, $1f_{5/2}$, and at higher energies, 
$1g_{9/2}$, valence neutron structures to states in a 
corresponding $_{20}$Ca core.  This is 
shown most directly by considering the cases of $^{50}$Sc and $^{52}$Sc, 
where this simple description explains many aspects of the known 
low-energy structures.  

\begin{figure*}[!tb]
\centering
\includegraphics[width = 0.85\textwidth]{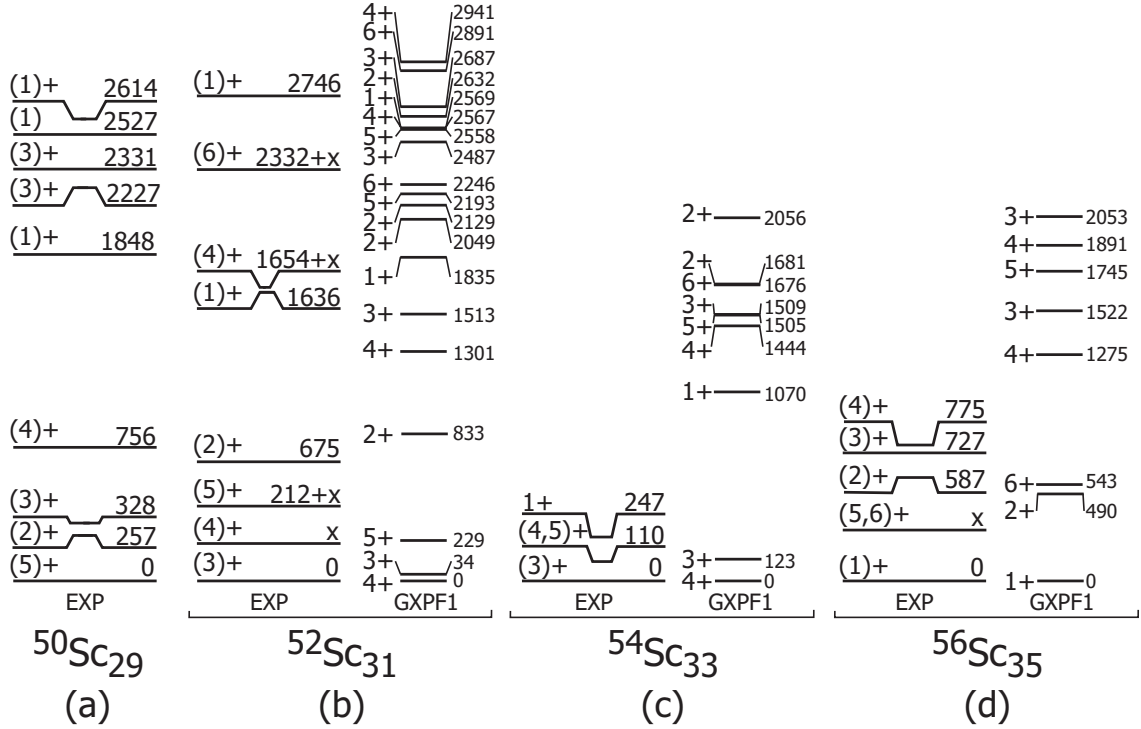}
\caption[Sc level schemes]{Experimentally determined levels below 3~MeV 
for the odd-odd, even-A (a) $^{50}$Sc, (b) $^{52}$Sc, (c) $^{54}$Sc and 
(d) $^{56}$Sc isotopes.  Level energies are labeled in keV.  
Theoretical predictions using the {\sc GXPF1} shell-model 
interaction are shown where available.}
\label{fig:sc_systematic}
\end{figure*}

$^{50}$Sc has one proton and one neutron particle outside 
of the doubly magic $^{48}$Ca core, and the lowest energy levels 
can be described in terms 
of the coupling of a valence $1f_{7/2}$ proton and 
$2p_{3/2}$ neutron.  The low-energy structure, known from 
$\beta$ decay~\cite{Albu84}, transfer reactions~\cite{Fist94}, 
and charge exchange reactions~\cite{Ajze85} can be seen in Fig.\ 
\ref{fig:sc_systematic}(a).  The four states lowest in energy arise 
from the configuration $(\pi$$1f_{7/2})^{1}$$\otimes$$(\nu$$2p_{3/2})^{1}$,
which produces four states ranging from $J^{\pi}$ = 2$^{+}$ to 5$^{+}$.  
The particle-particle coupling rules~\cite{Paar79} suggest that 
the states in this multiplet are arranged in a downward-opening 
parabola, in good agreement with the experimental energy levels.  
Promotion of the valence neutron in $^{50}$Sc to the $\nu$$2p_{1/2}$ 
level will give the configuration 
$(\pi$$1f_{7/2})^{1}$$\otimes$$(\nu$$2p_{1/2})^{1}$, producing two states with 
$J^{\pi}$=3$^{+}$, 4$^{+}$.  Two 3$^{+}$ states are identified in the level 
scheme of $^{50}$Sc above 2~MeV, and either may be a candidate for the 3$^{+}$ 
member of this doublet.  The energy spacing between multiplets arising 
from configurations involving the $\nu$$2p_{3/2}$ and $\nu$$2p_{1/2}$ states 
is directly related to the $\nu$$2p_{3/2}$-$\nu$$2p_{1/2}$ single-particle 
spacing, and thus the $\nu$$2p_{3/2}$-$\nu$$2p_{1/2}$ spin-orbit splitting.  
The apparent separation of these two multiplets in $^{50}$Sc is 
of order 2~MeV, a gap sufficient to account for the established 
$N$=32 subshell closure in $_{21}$Sc. 

The low-energy structure of $^{52}$Sc can be similarly 
described in terms of coupling of a valence $1f_{7/2}$ proton particle 
and $2p_{3/2}$ neutron hole, taking $^{52}$Ca as the inert core.  States in this odd-odd 
nucleus have been identified following $\beta$ decay~\cite{Huck85}, 
in-beam $\gamma$-ray spectroscopy following secondary fragmentation~\cite{Gade06_2} and 
deep-inelastic work~\cite{Forn08, Bhat09}.  The known levels are presented in 
Fig.\ \ref{fig:sc_systematic}(b).  As discussed in Refs.~\cite{Forn08, Bhat09}, 
the lowest-lying quartet of states 
in $^{52}$Sc can be associated with the configuration 
$(\pi$$1f_{7/2})^{1}$$\otimes$$(\nu$$2p_{3/2})^{-1}$.  The four states 
with $J^{\pi}$=2$^{+}$-5$^{+}$ should form an upward-opening parabola, since 
the coupling involves a proton particle and neutron hole.  Shell-model 
calculations based on the effective interactions 
GXPF1~\cite{Gade06} and GXPF1A~\cite{Forn08}, which assume an inert $^{48}$Ca 
core, both result in the near-degeneracy of the 3$^{+}$ and 4$^{+}$ states of this 
multiplet.  The energy separation between these states has not been 
established experimentally, but is likely small.  The success of the Pandya
transformation~\cite{Pand56} in relating the particle-particle 
$(\pi$$1f_{7/2})^{1}$$\otimes$$(\nu$$2p_{3/2})^{1}$ states of 
$^{50}$Sc to the particle-hole $(\pi$$1f_{7/2})^{1}$$\otimes$$(\nu$$2p_{3/2})^{-1}$ 
states in $^{52}$Sc also suggests that the energy separation between 
the 3$^{+}$ and 4$^{+}$ states is small in $^{52}$Sc, and that the 
configurations of these low-lying levels in both nuclei is fairly pure.  
Calculations using the GXPF1 effective interaction~\cite{Gade06} 
confirm the configurational purity of the four lowest-energy levels.  
The coupling of the $1f_{7/2}$ proton to an excited $2p_{1/2}$ 
neutron to give a pair of states with $J^{\pi}$=3$^{+}$, 4$^{+}$ 
accounts for the observed 4$^{+}$ state in $^{52}$Sc at $\sim$1.7~MeV.  
The 6$^{+}$ state at $\sim$2.3~MeV in $^{52}$Sc has been shown in 
shell-model calculations to have parentage including this configuration, 
though the extent of this contribution varies between shell-model 
interactions.  Calculations using the {\sc GXPF1A} 
interaction~\cite{Forn08} suggest that the 
primary parentage of the 2.3-MeV 6$^{+}$ state is 
$(\pi$$1f_{7/2})^{1}$$\otimes$$[\nu$$(2p_{3/2}^{2}2p_{1/2}^{1})]$, while the 
$(\pi$$1f_{7/2})^{1}$$\otimes$$[\nu$$(2p_{3/2}^{2}1f_{5/2}^{1})]$ 
configuration is predicted to dominate the 
first 8$^{+}$ state in $^{52}$Sc, experimentally observed at 
$\sim$3.6~MeV~\cite{Forn08}.  The energy separation 
between states arising from the $(\pi$$1f_{7/2})^{1}$$\otimes$$(\nu$$2p_{3/2})^{-1}$ and 
$(\pi$$1f_{7/2})^{1}$$\otimes$$[\nu$$(2p_{3/2}^{2}2p_{1/2}^{1})]$ configurations was 
$\sim$2~MeV, corresponding to the spin-orbit splitting between the 
$\nu$$2p_{3/2}$ and $\nu$$2p_{1/2}$ single-particle states, 
in agreement with the situation in $^{50}$Sc.  As noted by Fornal 
${\it et~al.}$~\cite{Forn08}, the apparent separation between the $\nu$$2p_{1/2}$ 
and $\nu$$1f_{5/2}$ single-particle states, as inferred from the energy spacing 
between the 6$^{+}$ and 8$^{+}$ states in $^{52}$Sc, is slightly smaller than 
that predicted in the {\sc GXPF1A} interaction.

The low-energy structure of $^{54}$Sc, with two additional neutrons, 
should exhibit significant change.  The $\nu$$2p_{3/2}$ orbital is now fully 
occupied, and the $^{54}$Sc structure at low energy should reflect 
coupling of the valence $1f_{7/2}$ proton particle to either a 
$2p_{1/2}$ or $1f_{5/2}$ neutron, taking $^{52}$Ca as 
the inert core.  The $\nu$$2p_{1/2}$ single-particle level is expected to lie 
below the $\nu$$1f_{5/2}$ state for the $_{21}$Sc isotopes, based on the 
observations in $^{52}$Sc.  Assuming such an ordering, the 
$(\pi$$1f_{7/2})^{1}$$\otimes$$(\nu$$2p_{1/2})^{1}$ 
coupling yields a doublet of states with $J^{\pi}$=3$^{+}$, 4$^{+}$, 
which is expected to be lowest in energy in $^{54}$Sc.  
Shell model results~\cite{Lidd04} predict the 3$^{+}$ and 4$^{+}$ 
states of this configuration to be close in energy, 
and thus the spin and parity of the $^{54}$Sc ground 
state from a theory perspective is not clear.  The ground state is tentatively
assigned $J^{\pi}$=3$^{+}$ in the present work.  Considering the situation 
in terms of the coupling, it may be natural to assume that the isomeric 
110-keV level corresponds to the 4$^{+}$ state arising from the 
$(\pi$$1f_{7/2})^{1}$$\otimes$$(\nu$$2p_{1/2})^{1}$ configuration.  However, 
within the simple coupling framework it is difficult 
to accept an in-multiplet M1 transition hindered so 
strongly as to yield the experimental isomer half-life of 2.8~$\mu$s.  
In this case, the two excited states identified in $^{54}$Sc, the isomeric (5)$^{+}$ level and higher-lying 
1$^{+}$ state, may be explained as arising from the configuration 
$(\pi$$1f_{7/2})^{1}$$\otimes$$(\nu$$1f_{5/2})^{1}$.  This coupling 
results in a sextet of 
states ranging from $J^{\pi}$=1$^{+}$ to 6$^{+}$.  The particle-particle 
coupling would produce a downward-facing parabola.  There are, however, inconsistencies 
in the experimental level ordering in $^{54}$Sc when compared with shell model calculations.  
The 1$^{+}$ state lies above the 5$^{+}$ state in 
energy in the level scheme as presented in Fig.\ \ref{fig:sc_systematic}(c), 
a result which is inconsistent with calculations using the GXPF1 
effective interaction.  An alternative $J^{\pi}$ assignment for the 
states in $^{54}$Sc with a 4$^{+}$ ground state and 6$^{+}$ 110-keV isomeric level is 
a possibility, but the placement of the 6$^{+}$ level below the 1$^{+}$ 
level is also unexpected.  The apparent 
experimental level ordering would be explained if the 
$(\pi$$1f_{7/2})^{1}$$\otimes$$(\nu$$2p_{3/2})^{-1}$ configuration was the 
origin of the 3$^{+}$ and 5$^{+}$ states.  However, as the 
$\nu$$2p_{3/2}$-$\nu$$2p_{1/2}$ spin-orbit splitting is a fairly constant 
$\sim$2~MeV, this multiplet is expected at much higher excitation 
energy in $^{54}$Sc, making this possibility unlikely.  

An alternative to the 5$^{+}$ assignment for the 110-keV isomeric state 
is a 4$^{+}$ assignment.  While not expected based on comparison of the 
deduced isomer lifetime with Weisskopf single-particle estimates, 
which favor an E2 multipolarity for the 110-keV transition, a 4$^{+}$ 
spin and parity cannot be excluded for the 110-keV state.  Calculations 
using the GXPF1~\cite{Honm02}, GXPF1A~\cite{Honm05} and KB3G~\cite{Pove01} 
interactions all predict a doublet of states near the $^{54}$Sc ground 
state with $J^{\pi}$=3$^{+}$ and 4$^{+}$.  The transition probabilities, 
$B(M1: 3^{+}\to 4^{+})$ and $B(E2: 3^{+}\to 4^{+})$ calculated with these 
three interactions are shown in Table~\ref{tab:54Sc110Transitions}, along 
with the calculated partial half-lives.  In all cases, the M1 component 
would dominate the transition.  However, the $B(M1: 3^{+}\to 4^{+})$ values 
are small, giving rise to half-lives of order nanoseconds or longer -- 
much slower than the picoseconds expectation from the single-particle 
Weisskopf estimate.  The $B(M1: 3^{+}\to 4^{+})$ value is very sensitive 
to the degree of mixing in the wavefunctions for the 3$^{+}$ and 
4$^{+}$ states.  With increased configuration mixing, 
there is increased suppression of the $B(M1: 3^{+}\to 4^{+})$ value, 
hindering the transition.  A 4$^{+}$ spin and parity assignment is in 
better agreement with the expectation for the level ordering in $^{54}$Sc.  
However, more work is required to make a firm spin assignment for the 110-keV 
isomeric state in $^{54}$Sc.

\begin{table*}
\centering
\caption[Calculated transition probabilities for the 3$^{+}_{1}$ $\to$ 4$^{+}_{1}$ transition in $^{54}$Sc]{Calculated transition probabilities and half-lives for the transition between the 3$^{+}$ and 4$^{+}$ states in $^{54}$Sc, using the GXPF1, GXPF1A and KB3G shell-model effective interactions.  The Weisskopf single-particle estimates for the half-life of a 110-keV transition are also included.}
\label{tab:54Sc110Transitions}
\begin{ruledtabular}
\begin{tabular}{ccccc}
 & GXPF1 & GXPF1A & KB3G & Single-Particle \\
\hline
$B(M1: 3^{+}\to 4^{+})$ ($\mu_{N}^{2}$) & 0.02611 &  0.00369 & 0.00080 & \\
$T_{1/2}(M1)$ (s) & 8.10$\times$10$^{-10}$ & 1.56$\times$10$^{-8}$ & 8.94$\times$10$^{-7}$ & 9.30$\times$10$^{-12}$ \\
$B(E2: 3^{+}\to 4^{+})$ ($e^{2}fm^{4}$) & 5.07 & 6.24 & 6.96 & \\
$T_{1/2}(E2)$ (s) & 4.05$\times$10$^{-6}$ & 1.72$\times$10$^{-5}$ & 1.03$\times$10$^{-3}$ & 2.89$\times$10$^{-6}$ \\
\end{tabular}
\end{ruledtabular}
\end{table*}

The 1$^{+}$ state at energy 247 keV in $^{54}$Sc can only be explained by the 
$(\pi$$1f_{7/2})^{1}$$\otimes$$(\nu$$1f_{5/2})^{1}$ configuration.  Even if 
this is the only member of the multiplet identified, the energy 
separation between the ground state with 
$(\pi$$1f_{7/2})^{1}$$\otimes$$(\nu$$2p_{1/2})^{1}$ and the excited states from
$(\pi$$1f_{7/2})^{1}$$\otimes$$(\nu$$1f_{5/2})^{1}$ appears to be small.  Shell 
model results using the GXPF1 effective interaction~\cite{Lidd04} predict a 
$\nu$$2p_{1/2}$-$\nu$$1f_{5/2}$ separation of more than 1~MeV, which manifests 
itself in $^{54}$Sc as an expanded low-energy level structure [see Fig.~\ref{fig:sc_systematic}(c)].  The compression 
observed here between multiplets arising from coupling with the 
$\nu$$2p_{1/2}$ and $\nu$$1f_{5/2}$ neutron states suggests a 
$\nu$$2p_{1/2}$-$\nu$$1f_{5/2}$ single-particle energy separation 
smaller than that assumed in the GXPF1 Hamiltonian, and is 
possibly inconsistent with a robust $N$=34 subshell closure.  

Another rapid structure change should be evident in $^{56}$Sc.  The 
addition of two neutrons to $^{54}$Sc should fill the $\nu$$2p_{1/2}$ 
single-particle orbital.  The unpaired valence neutron will then occupy the 
$1f_{5/2}$ state.  Under this assumption, the lowest energy 
states in $^{56}$Sc are expected to arise from the 
$(\pi$$1f_{7/2})^{1}$$\otimes$$(\nu$$1f_{5/2})^{1}$ configuration.  
The resulting sextet 
of states with $J^{\pi}$=1$^{+}$ to 6$^{+}$ should be arranged in a 
downward facing parabola, with the 4$^{+}$ state at the vertex.  The challenge
in comparing expectations in $^{56}$Sc to the new data is the unknown position 
of the higher-spin $\beta$-decaying state, relative to the low-energy 
structure established by the isomer decay.  However, the states at low 
energy in $^{56}$Sc with $J^{\pi}$=1$^{+}$, 2$^{+}$ and (5,6)$^{+}$ are difficult 
to explain in a simple way outside of the 
$(\pi$$1f_{7/2})^{1}$$\otimes$$(\nu$$1f_{5/2})^{1}$  configuration.  
The 3$^{+}$ and 4$^{+}$ 
states cannot be trivially assigned the same configuration, as they may also 
arise from the $(\pi$$1f_{7/2})^{1}$$\otimes$$(\nu$$2p_{1/2})^{1}$ coupling, 
depending on the relative position of the 
$\nu$$2p_{1/2}$ and $\nu$$1f_{5/2}$ single-particle orbitals.  In fact, 
significant mixing would be expected between the 3$^{+}$/4$^{+}$ states 
in the $\pi$$1f_{7/2}$-$\nu$$1f_{5/2}$ and $\pi$$1f_{7/2}$-$\nu$$2p_{1/2}$ 
multiplets.  This mixing is already reflected in the GXPF1 calculated levels, 
even with a large $\nu$$2p_{1/2}$-$\nu$$1f_{5/2}$ gap.

\subsection{\label{sec4:level2c}Systematic variation of the Ti levels}

\begin{figure}[!tb]
\centering
\includegraphics[width = 0.45\textwidth]{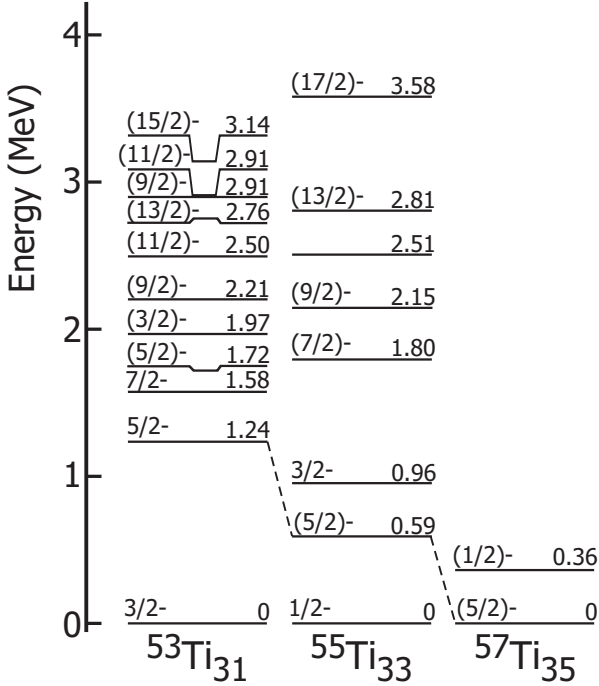}
\caption[Ti level schemes]{Experimentally-determined levels below 4~MeV 
for odd-$A$ $^{53,55,57}$Ti isotopes.}
\label{fig:ti_systematic}
\end{figure}

The experimentally-known levels below 4~MeV in the odd-$A$ neutron-rich 
$^{53,55,57}_{\phantom{53,55,}22}$Ti isotopes are represented in 
Fig.\ \ref{fig:ti_systematic}.  The levels of the $_{22}$Ti isotopes have 
been well characterized~\cite{Jans02, Dinc05, Forn05, Forn04, Zhu07} by the 
GXPF1A~\cite{Honm05} shell-model effective interaction.  

Within the {\sc GXPF1A} interaction, the 
ordering of the neutron single-particle states places the $2p_{1/2}$ 
orbital below the $1f_{5/2}$ orbital for the Ti isotopes.  Given 
such an ordering, $^{55}$Ti$_{33}$ is expected to have a 1/2$^{-}$ 
ground state, as the first neutron above the $N$=32 subshell 
closure should occupy the $2p_{1/2}$ orbital.  The first 
excited state in $^{55}$Ti then corresponds 
to excitation of the valence neutron into the $1f_{5/2}$ orbital.  
The recent direct determination of the ground state spin and parity in 
$^{55}$Ti~\cite{Maie09} provides confirmation for the ordering 
of the $\nu$$2p_{1/2}$ and $\nu$$1f_{5/2}$ orbitals, 
with the $\nu$$2p_{1/2}$ orbital lying lower in energy, similar to 
the ordering in the $_{21}$Sc isotopes.  Thus, it appears that the 
$\nu$$2p_{1/2}$-$\nu$$1f_{5/2}$ level ordering changes already 
in going from $_{23}$V to $_{22}$Ti, though a significant energy 
gap between these orbitals is not present in the Ti isotopes~\cite{Zhu07}. 

Based on the assumed neutron-level ordering, $^{57}$Ti$_{35}$ is expected to 
have a 5/2$^{-}$ ground state, with a neutron configuration of 
$2p_{3/2}^{4}2p_{1/2}^{2}1f_{5/2}^{1}$, and a first excited state 
with $J^{\pi}$=1/2$^{-}$, corresponding to excitation of 
a $2p_{1/2}$ neutron into the $1f_{5/2}$ orbital.  Thus, 
this is essentially an inversion of the two lowest states in $^{55}$Ti.  
Calculations with the {\sc GXPF1A} interaction have placed this first 
excited state at an energy of 422~keV~\cite{Lidd05}.  In the present results, 
one state has been placed in the level scheme of $^{57}$Ti at an energy of 
364~keV, which may correspond to the 1/2$^{-}$ first excited state.  If this 
is the case, $\beta$ decay from the $^{57}$Sc 7/2$^{-}$ ground state should 
not directly populate this state, and it must be fed indirectly 
from higher-lying states.  However, the 364-keV transition is much 
stronger than the other transitions observed in the decay, and direct 
feeding cannot be excluded.  Allowed $\beta$ decay from $^{57}$Sc, 
with a 7/2$^{-}$ ground state, populating the first-excited state in 
$^{57}$Ti would suggest a reversal of the level ordering shown in 
Fig.\ \ref{fig:ti_systematic}, and would present a 
significant challenge for theory.
However, before the origin of the 364-keV state can be confirmed, a more 
complete picture of the low-energy states in $^{57}$Ti is required.

%%%%%%%%%%%%%%%%%%%%%%  CONCLUSION   %%%%%%%%%%%%%%%%%%%%%%%%%%%%%%%%%%%%

\section{\label{sec5:level1}Summary}

The low-energy levels of neutron-rich $^{53,54,56}$Sc and $^{53-57}$Ti 
have been investigated through the $\beta$ decay of the parent nuclides 
$^{53,54}$Ca and $^{53-57}$Sc, as well as through prompt isomeric $\gamma$-ray 
emission from $^{54,56}$Sc.  The low-energy structures of the Sc isotopes 
can be considered in the extreme single-particle model as the result of 
coupling of the valence $1f_{7/2}$ proton to states in the 
corresponding $_{20}$Ca core.  Such a description works well in $^{53}$Sc, 
where the valence proton weakly couples to states in doubly-magic 
$^{52}_{20}$Ca$_{32}$.  The success of this description provides further 
evidence for the validity of the $N$=32 subshell closure in the Ca isotopes.  
Interpretation of the low-energy levels in the even-A Sc isotopes as 
resulting from the simple coupling of a $1f_{7/2}$ proton and valence 
neutrons to a Ca core provides preliminary information regarding the 
energy separation between neutron single-particle levels.  While more work 
is required to complete the relevant multiplets of states, early indications 
suggest that the low-energy levels in the Sc isotopes are compressed, and that 
a significant $N$=34 subshell gap does not exist between the $2p_{1/2}$ and 
$1f_{5/2}$ neutron levels in the $_{21}$Sc isotopes.  More 
information is required in the $_{20}$Ca isotopes to determine 
conclusively whether the removal of the final $1f_{7/2}$ proton 
is sufficient to create a substantial subshell closure at $N$=34.   

\begin{acknowledgments}
The authors thank the NSCL operations staff for providing the primary 
and secondary beams for this experiment and NSCL $\gamma$ group for 
assistance setting up the Ge detectors from SeGA.  This work was 
supported in part by the National Science Foundation under Grant No. 
PHY-06-06007 and by the U.S. Department of Energy, Office of Nuclear 
Physics, under contracts No. DE-AC02-06CH11357 (ANL) and 
DEFG02-94ER40834 (University of Maryland) and by the Polish Scientific 
Committee grant 1PO3B 059 29.  HLC and GFG would like to acknowledge 
support from the Natural Science and Engineering Research 
Council (NSERC) of Canada.
\end{acknowledgments}

\end{document}